\documentclass[a4paper,fleqn,usenatbib]{mnras}

\usepackage{mathptmx}

\usepackage[T1]{fontenc}
\usepackage{ae,aecompl}

\usepackage{graphicx}
\usepackage{amsmath}	
\usepackage{amssymb}	
\usepackage{mathrsfs}
\usepackage{url}
\usepackage{bm}
\usepackage{booktabs}
\usepackage{multicol}
\usepackage{comment}

\newcommand{\msun}{\ensuremath{\,\textrm{M}_{\odot}}}

\title[The progenitors of compact-object binaries]{The progenitors of compact-object binaries: impact of metallicity, common envelope and natal kicks} 

\author[Giacobbo \& Mapelli]{
Nicola Giacobbo,$^{1,2,3,4}$\thanks{E-mail: giacobbo.nicola@gmail.com}
Michela Mapelli,$^{2,3,4}$
\\
$^{1}$Dipartimento di Fisica e Astronomia ``G. Galilei'',
    Universit\`a di Padova, vicolo dell'Osservatorio 3, I-35122 \\
$^{2}$INAF, Osservatorio Astronomico di Padova, vicolo dell'Osservatorio 5, I--35122 Padova, Italy\\
$^{3}$Institute for Astrophysics and Particle Physics, University of Innsbruck, A--6020, Innsbruck, Austria\\ 
$^{4}$INFN, Milano Bicocca, Piazza della Scienza 3, I--20126, Milano, Italy\\
}

\date{Accepted XXX. Received YYY; in original form ZZZ}

\pubyear{2017}

\begin{document}
\label{firstpage}
\pagerange{\pageref{firstpage}--\pageref{lastpage}}
\maketitle

\begin{abstract}
Six gravitational wave events have been reported by the LIGO-Virgo collaboration (LVC), five of them associated with black hole binary (BHB) mergers and one with a double neutron star (DNS) merger, while the coalescence of a black hole-neutron star (BHNS) binary is still missing. We investigate the progenitors of double compact object binaries with our population-synthesis code {\sc MOBSE}. {\sc MOBSE} includes advanced prescriptions for mass loss by stellar winds (depending on metallicity and on the Eddington ratio) and a formalism for core-collapse, electron-capture and (pulsational) pair instability supernovae. We investigate the impact of progenitor's metallicity, of the common-envelope parameter $\alpha{}$ and of the natal kicks on the properties of DNSs, BHNSs and BHBs. We find that neutron-star (NS) masses in DNSs span from 1.1 to 2.0 M$_\odot$, with a preference for light NSs, while NSs in merging BHNSs have mostly large masses ($1.3-2.0$ M$_\odot$). BHs in merging BHNSs are preferentially low mass ($5-15$ M$_\odot$). BH masses in merging BHBs strongly depend on the progenitor's metallicity and span from $\sim{}5$ to $\sim{}45$ M$_\odot$. The local merger rate density of both BHNSs and BHBs derived from our simulations is consistent with the values reported by the LVC in all our simulations. In contrast, the local merger rate density of DNSs matches the value inferred from the LVC only if low natal kicks are assumed. This result adds another piece to the intricate puzzle of natal kicks and DNS formation.

\end{abstract}




\section{Introduction}
On August 17 2017, the LIGO-Virgo collaboration (LVC, \citealt{LIGOdetector,Virgodetector}) captured the first gravitational wave (GW) signal from a double neutron star (DNS) merger \citep{Abbott2017d,Abbott2017e}. GW170817 was accompanied by electromagnetic radiation over a large range of wavelengths, from radio to gamma-rays \citep{Abbott2017z,Goldstein2017,Savchenko2017,Margutti2017,Coulter2017,Soares-Santos2017,Chornock2017,Cowperthwaite2017,Nicholl2017,Pian2017,Alexander2017}, marking the beginning of multi-messenger astronomy. Besides GW170817, five other GW detections were reported so far (GW150914, GW151226, GW170104, GW170608, GW170814), all of them interpreted as the merger of two black holes (BHs, \citealt{Abbott2016a,Abbott2016b,Abbott2016c,Abbott2016d,Abbott2017a,Abbott2017b,Abbott2017c}).

Unlike BH binaries (BHBs), whose very existence was revealed by direct detections of GWs \citep{Abbott2016a}, DNSs were observed well before GW170817: PSR~$B1913+16$ was discovered already in 1974 \citep{Hulse1975}, followed by about a dozen similar binaries \citep{Tauris2017}, including a double pulsar \citep{Burgay2003,Lyne2004}. Together with GW170817, these highly relativistic systems give us a unique grasp on the behaviour of matter under extreme conditions.

Now that both BHBs and DNSs have been detected by the LVC, the coalescence of a neutron star (NS) with a BH is the only missing merger event that we expect to observe in the frequency range of ground-based GW detectors. No BH-NS binaries (BHNSs) have been discovered so far by radio surveys. 

Previous work investigates the formation of DNSs, BHNSs and BHBs both from isolated binaries (e.g. \citealt{Tutukov1973,Flannery1975, Bethe1998,Belczynski2002,Voss2003, Dewi2003, Podsiadlowski2004,Podsiadlowski2005,Dewi2005,Tauris2006,Portegies1998,  Portegies2000, Belczynski2007,  Bogomazov2007,  Dominik2012, Dominik2013,  Dominik2015,  Mennekens2014, Tauris2015, Tauris2017, Demink2015,DeMink2016,Marchant2016,Chruslinska2018,Mapelli2017,Giacobbo2018,Kruckow2018,Shao2018}) and from dynamics (e.g. \citealt{Kulkarni1993,Sigurdsson1993a,Sigurdsson1993b,Sigurdsson1995,Phinney1996,Colpi2003,Mapelli2005,Grindlay2006,Ivanova2008, Clausen2013,Mapelli2014,Ziosi2014,Rodriguez2015,Rodriguez2016,Mapelli2016,Askar2016,Antonini2017,Petrovich2017,Banerjee2017,Belczynski2018}).

Despite this effort, the evolution of compact-object binaries is still highly uncertain. The merger rate density of DNSs inferred from GW170817 is $R_{\rm DNS}=1540^{+3200}_{-1220}$ Gpc$^{-3}$ yr$^{-1}$, consistent with recent estimates from short gamma-ray bursts ($\sim{}8-1800$ Gpc$^{-3}$ yr$^{-1}$ according to \citealt{coward2012}; $\sim{}500-1500$  Gpc$^{-3}$ yr$^{-1}$ based on the analysis of \citealt{petrillo2013}; $\sim{}92-1154$  Gpc$^{-3}$ yr$^{-1}$ as estimated by \citealt{siellez2014}, and $270^{+1580}_{-180}$  Gpc$^{-3}$ yr$^{-1}$ according to \citealt{fong2015}), but quite large with respect to the rate predicted from kilonovae  ($\sim{}63^{+63}_{-31}$  Gpc$^{-3}$ yr$^{-1}$, \citealt{jin2015}). Simulations of isolated and dynamically formed DNSs show that it is very difficult to match such a high rate, unless rather extreme assumptions about common envelope (e.g. \citealt{Chruslinska2018}) or natal kicks (e.g. \citealt{Giacobbo2018b,Mapelli2018}) are made.

In this paper, we use our population-synthesis code {\sc MOBSE} \citep{Giacobbo2018} to investigate the formation of DNSs, BHNSs and BHBs from isolated binaries  and to analyze the effect of natal kicks and common-envelope efficiency on the merger rate of DNSs, BHBs and BHNSs. In its current version, {\sc MOBSE} includes up-to-date prescriptions for stellar winds (accounting for the metallicity and for the Eddington ratio of the progenitor star), for core-collapse supernovae (SNe), electron-capture SNe, pulsational pair-instability SNe and pair-instability SNe.

\section{Methods}
\label{sec:2}
{\sc MOBSE}, which stands for ``massive objects in binary stellar evolution'', is an updated version of the populations synthesis code {\sc BSE} \citep{Hurley2000,Hurley2002}. {\sc MOBSE} is meant to improve the treatment of massive stars and stellar remnants. We refer to \cite{Giacobbo2018} and \cite{Mapelli2017} for a detailed description of {\sc MOBSE}. Here, we just summarize the main features of {\sc MOBSE} with respect to {\sc BSE}.

\subsection{Single star evolution and SNe}
{\sc MOBSE} includes a new treatment of stellar winds for hot massive stars, based on \cite{Vink2001} for O-type and B-type stars, on \cite{VinkdeKoter2005} for Wolf-Rayet stars, and on \cite{Belczynski2010} for luminous blue variable stars. For all types of hot massive stars, we adopt a description of the mass loss as $\dot{M}\propto{}Z^{\beta{}}$, where $Z$ is the metallicity and $\beta{}=0.85$ if the Eddington ratio $\Gamma{}<2/3$, $\beta{}=2.45-2.4\,{}\Gamma{}$ if $2/3\leq{}\Gamma{}<{}1$, and $\beta{}=0.05$ if $\Gamma{}\ge1$ \citep{Chen2015}.

We have also updated the prescriptions for core radii following \cite{hall2014} and we have extended the mass range up to 150 M$_\odot$ \citep{Mapelli2016}.

\begin{center}
	\begin{figure}
		\includegraphics[width=8.5cm]{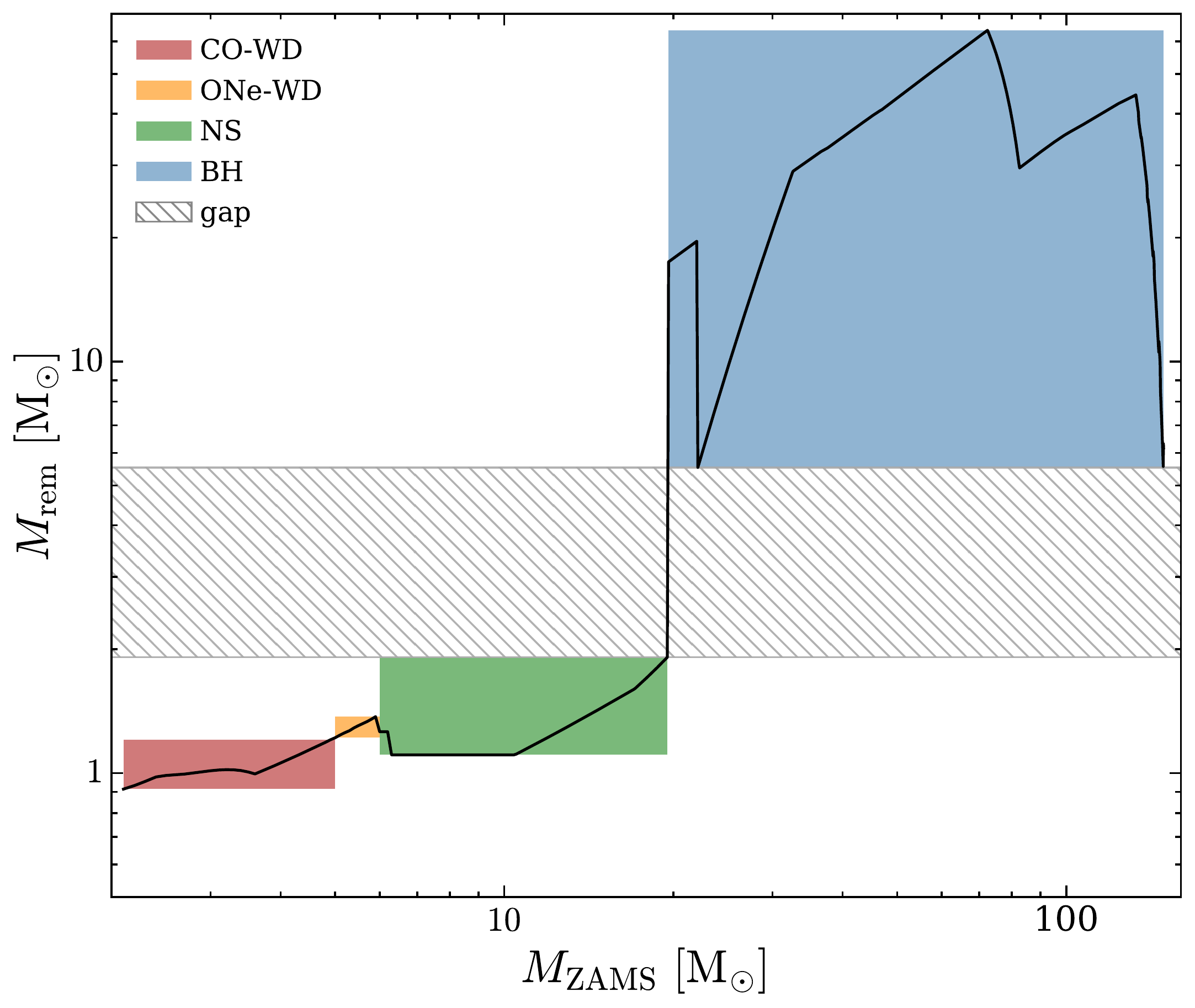}
		\caption{\label{fig:spec}Mass spectrum of compact remnants as a function of the zero-age main sequence (ZAMS) mass of the progenitor stars. The coloured regions (red, orange, green and blue) identify the mass ranges of different compact objects (Carbon-Oxygen WDs, Oxygen-Neon WDs, NSs and BHs, respectively). The hashed region shows the mass gap between NSs and BHs predicted by the rapid core-collapse SN model \citep{Fryer2012}. The metallicity is $Z=0.0002$.}
	\end{figure}	
\end{center}

The treatment of SNe was also deeply revised: we have included pulsational pair instability and pair instability SNe as described in \cite{Spera2017}, and we have added two new prescriptions for iron core-collapse SNe: the delayed and the rapid models presented in \cite{Fryer2012} (see also \citealt{Spera2015}). In this paper, we always adopt the rapid core-collapse SN mechanism, because it allows us to reproduce the mass gap of compact objects between $\sim{}2$ and $\sim{}5$ M$_\odot$ \citep{Ozel2010,Farr2011}, as it can be seen from Figure~\ref{fig:spec}.

We have also updated the treatment for electron-capture SNe. In the case of both an electron-capture SN and an accretion-induced white dwarf (WD) collapse, the NS forms when the degenerate Oxygen-Neon (ONe) core collapses as a consequence of electron-capture reactions, inducing a thermonuclear runaway.  In {\sc MOBSE}, we assume that if a growing degenerate ONe core reaches the mass $M_{\rm ECSN}=1.38$ \msun~it collapses due to the electron-capture on $^{24}$Mg and on $^{20}$Ne \citep{Miyaji1980,  Nomoto1984, Nomoto1987, Nomoto1991, Kitaura2006, Fisher2010, Jones2013, Takahashi2013, Schwab2015, Jones2016}, otherwise it forms a ONe WD, which can still collapse to a NS if it will accrete sufficient mass.

 The outcome of the electron-capture collapse is a NS. We compute the final mass of the newly-born NS by using the formula suggested by \citet{Timmes1996}
\begin{equation}
	M_{\rm rem,grav} = \frac{\sqrt{1 + 0.3M_{\rm rem,bar}}-1}{0.15}=1.26 \msun~,
\end{equation}
where $M_{\rm rem, bar} = M_{\rm ECSN}$ is the baryonic mass of the NS including neutrinos, and $M_{\rm rem,grav}$ is the final mass of the NS considering the mass loss due to neutrinos emission.

The current version of {\sc MOBSE} does not include any specific treatment for the initial spin of BHs and NSs.
What drives the initial spin magnitude and direction of a compact object is still poorly understood and constrained. We refer to \cite{Postnov2017}, \cite{Wysocki2018}, \cite{Belczynski2017spin} and \cite{Arcasedda2018} for a more detailed discussion of BH spins.

\subsection{SN kicks}
\label{sec:kicks}
In {\sc MOBSE}, the natal kick of a NS is drawn from a Maxwellian velocity distribution,
\begin{equation}
	f(v,\sigma)=\sqrt{\frac{2}{\pi}}\frac{v^2}{\sigma^3}\exp\left[{-\frac{v^2}{2\sigma^2}}\right] \qquad v \in~ [0,\infty{})
\end{equation}
where $\sigma$ is the one dimensional root-mean-square.

Here, we have implemented in {\sc MOBSE} the possibility to draw the natal kick from two Maxwellian curves with a different root-mean-square: $\sigma_{\rm CCSN}$ and $\sigma_{\rm ECSN}$, for core-collapse and electron-capture SNe, respectively.

For electron-capture SNe, we draw the kicks from a Maxwellian with one-dimensional root-mean square (1D rms) $\sigma{}_{\rm ECSN}=15$ km s$^{-1}$, corresponding to an average velocity of $\sim{}23$ km s$^{-1}$ \citep{Giacobbo2018b}. This approach is justified by the fact that  an electron-capture collapse is a more rapid process with respect to the iron core-collapse explosion and for that reason the asymmetries are more difficult to develop \citep{Kitaura2006,Dessart2006,Janka2012}. The consequence is that the newborn NS has a low natal kick \citep{Dessart2006,Jones2013,Schwab2015}.

For iron core-collapse SNe the situation is more puzzling. \cite{Hobbs2005} report the proper motion of 233 Galactic single pulsars and describe the kick velocities using a Maxwellian with 1D rms $\sigma{}_{\rm CCSN}=265$ km s$^{-1}$, corresponding to an average velocity of $\sim{}420$ km s$^{-1}$. Several studies \citep{Cordes1998,Arzoumanian2002,Brisken2003,Schwab2010,Verbunt2017} claim that the velocity distribution proposed by \citet{Hobbs2005} underestimates the number of pulsars with a low velocity and suggest that the natal kick distribution of NSs is better represented by a bimodal velocity distribution. For instance, two out of nine accurate pulsar velocities computed by \citet{Brisken2002} are smaller than $40$ km s$^{-1}$. Moreover, \citet{Pfahl2002} study a new class of high-mass X-ray binaries with small eccentricities and long orbital periods, which imply a low natal kick velocity ($\lesssim 50$ km s$^{-1}$) for the newborn NSs. This bimodal distribution might result from two different mechanisms of NS formation \citep{Heuvel2007,Beniamini2016}.

Finally \cite{Tauris2017} suggest that not only electron-capture SNe, but even iron core-collapse SNe could be associated with low kicks $\lesssim{}50$ km s$^{-1}$, when the SN is {\it ultra-stripped} \citep{Tauris2013,Tauris2015, Bray2016}. A SN is ultra-stripped if the exploding  star is member of a binary system and was heavily stripped by its companion. In this case, the mass ejected during the SN is small ($\lesssim{}0.1$ M$_\odot$) and thus the kick is also small ($<50$ km s$^{-1}$, e.g. \citealt{Suwa2015}). This suggests that the natal kicks of single NSs could be significantly different from the kicks of NSs born in close binary systems.

Given this uncertainty, we decided to simulate two extreme cases for the kick of iron core-collapse SNe. In the first case (hereafter: high-velocity kicks), we draw core-collapse SN kicks from a Maxwellian distribution with 1D rms $\sigma{}_{\rm CCSN}=265$ km s$^{-1}$, as derived by \cite{Hobbs2005}. In the second case (hereafter: low-velocity kicks), we draw core-collapse SN kicks from a Maxwellian distribution with 1D rms $\sigma{}_{\rm CCSN}=15$ km s$^{-1}$, i.e. the same as for electron-capture SNe (see Table~\ref{tab:sims}).

Finally, while NSs receive the full kick drawn from the Maxwellian distribution, BHs receive a natal kick that is reduced by the amount of fallback as $v_{\rm BH}=(1-f_{\rm fb})\,{}v$, where $f_{\rm fb}$ is the fallback parameter and $v$ is the velocity randomly sampled from the Maxwellian curve \citep{Fryer2012}.
\subsection{Binary evolution}
The description of binary evolution is the same as in \cite{Hurley2002}, apart from several features of the common envelope (CE) phase.

To describe the CE phase, we adopt the $\alpha{}\lambda$ formalism \citep[see][]{Webbink1984,Ivanova2013}, where $\alpha{}$ quantifies the energy available to unbind the envelope, and $\lambda{}$ measures the concentration of the envelope. In our simulations, $\lambda$ depends on the stellar type (i.e. mass and luminosity) to account for the contribution of recombinations. To compute $\lambda$ we used the prescriptions derived by \citet{Claeys2014} (see their Appendix A for more details) which are based on \citet{Dewi2000}.

In contrast, $\alpha{}$ is a free parameter. In this paper, we assume $\alpha{}=1,\,{}3$ and 5 (see Table~\ref{tab:sims}).

With respect to \cite{Hurley2002},  we have revised the treatment of Hertzsprung-gap (HG) donors in CE: HG donors are assumed to always merge with their companions if they enter a CE phase. This is justified by the fact that HG stars have not yet developed a steep density gradient between core and envelope, and allows us to match the merger rate of BHBs \citep{Mapelli2017} inferred from  LIGO-Virgo observations \citep{Abbott2016d}.


\begin{table}
	\begin{center}
		\caption{Initial conditions of the {\sc MOBSE} simulations. \label{tab:sims}}
		\begin{tabular}{ccccc}
			\toprule
			{\bf ID} & $\bm{\sigma_{\rm{ECSN}}}$ & $\bm{\sigma_{\rm{CCSN}}}$  &  {\bf SN} &  $\bm{\alpha }$  \\
			\midrule
			$\alpha1$ & 15.0 km/s & 265.0 km/s & rapid & 1.0 \\
			$\alpha3$ & 15.0 km/s & 265.0 km/s & rapid & 3.0 \\
			$\alpha5$ & 15.0 km/s & 265.0 km/s & rapid & 5.0 \\
			CC15$\alpha1$ & 15.0 km/s & 15.0 km/s & rapid & 1.0 \\ 
			CC15$\alpha3$ & 15.0 km/s & 15.0 km/s & rapid & 3.0 \\ 
			CC15$\alpha5$ & 15.0 km/s & 15.0 km/s & rapid & 5.0 \\ 
			\bottomrule	
		\end{tabular}
	\end{center}
	{\small Column 1: simulation name; columns 2 and 3: 1D rms of the Maxwellian distribution for electron-capture SN kicks and for iron core-collapse SN kicks (see sec.~\ref{sec:kicks}); column 4: SN model \citep{Fryer2012,Giacobbo2018}; column 5: values of $\alpha$ adopted in the CE formalism. 
	}
	
\end{table}

\begin{center}
	\begin{figure*}
		\includegraphics[width=17cm]{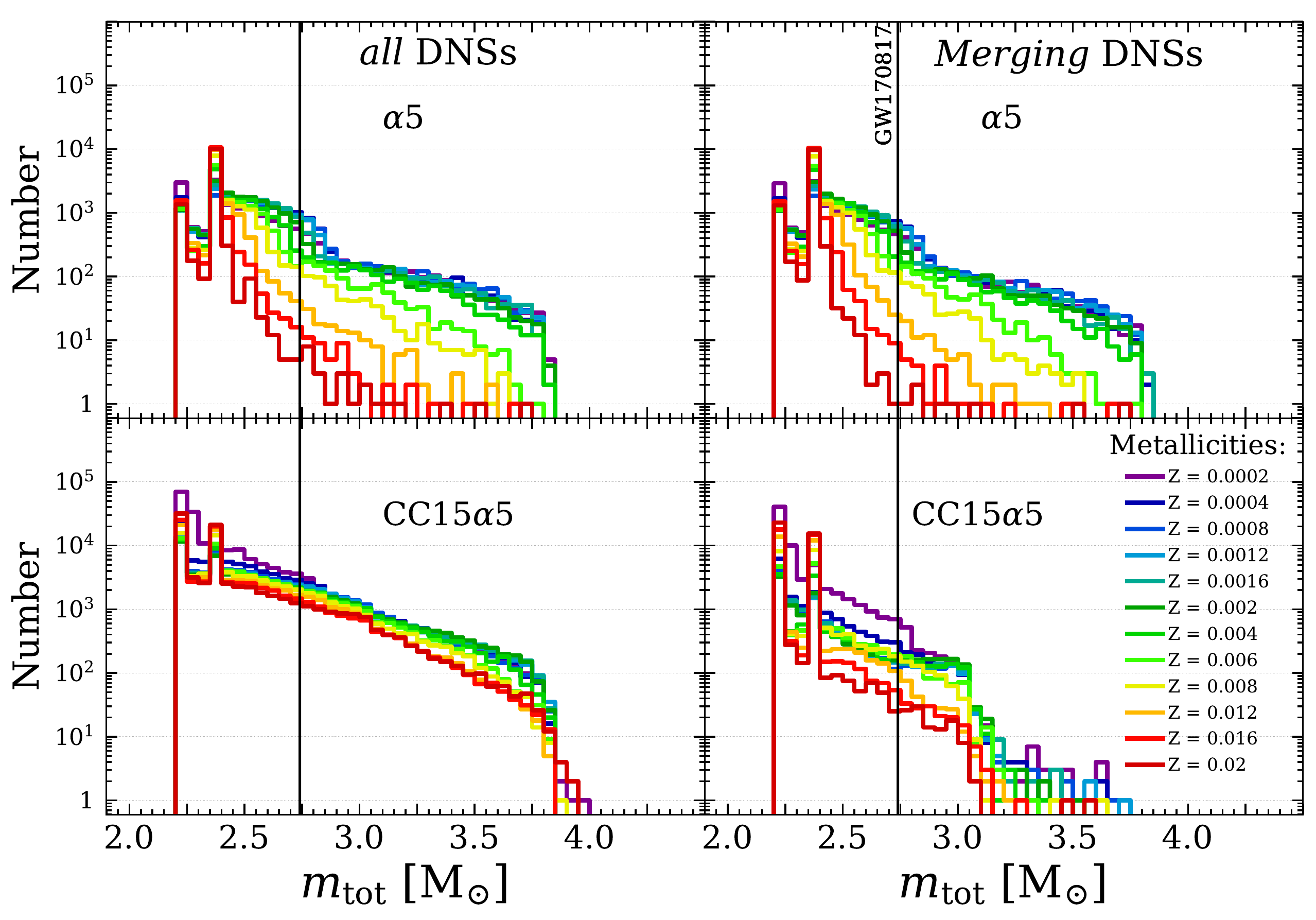}
		\caption{\label{fig:mtot_DNS}Distribution of total masses of DNSs in simulations $\alpha{}5$ (top panels) and CC15$\alpha{}5$ (bottom panels). Left-hand panels: all DNSs that formed in our simulations are shown. Right-hand panels: only DNSs merging in less than a Hubble time are shown.  The vertical lines are the total mass of the GW event GW170817~\citep{Abbott2017d}. The bin width is $0.05$ \msun.}
	\end{figure*}	
\end{center}
\begin{center}
	\begin{figure*}
		\includegraphics[width=14cm]{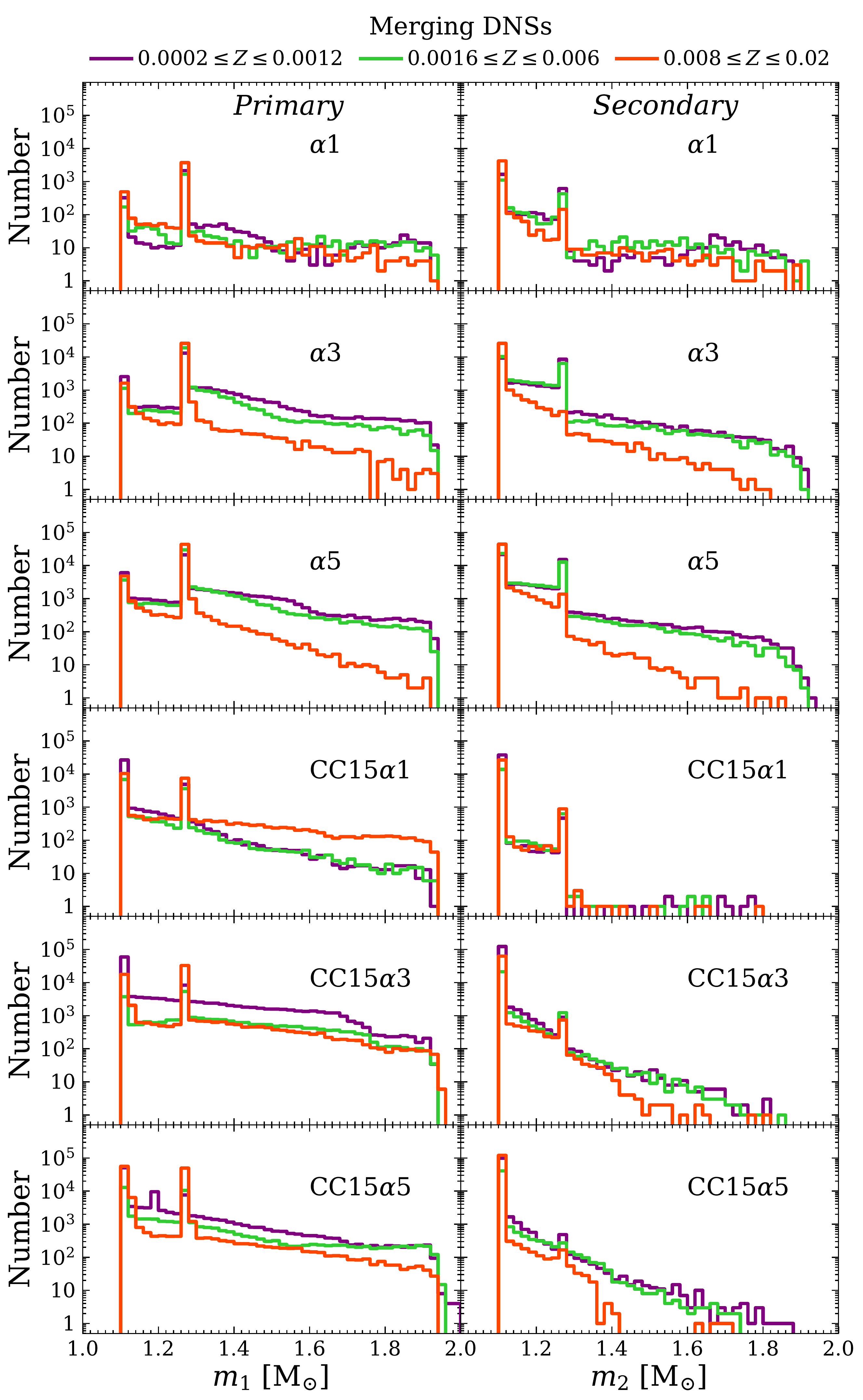}
		\caption{\label{fig:m12DNS}Primary masses ($m_1$, left-hand panels) and secondary masses ($m_2$, right-hand panels) of DNSs merging within a Hubble time in all simulations presented in this paper. Violet histogram: merging DNSs with metallicity  $0.0002\leq{}Z\leq{}0.0012$; green histogram: merging DNSs with metallicity $0.0016\leq{}Z\leq{}0.006$; red histogram: merging DNSs with metallicity $0.008\leq{}Z\leq{}0.02$. The bin width for both $m_1$ and $m_2$ is $0.02$ \msun.}
	\end{figure*}	
\end{center}

\subsection{Simulations and initial distributions}
\label{sec:2.3}
Here we describe the initial conditions used to perform our population-synthesis simulations.   
We randomly draw the mass of the primary star ($m_{\mathrm{1}}$) from a Kroupa initial mass function \citep[IMF,][]{Kroupa2001}
\begin{equation}
	\mathfrak{F}(m_1) ~\propto~ m_1^{-2.3} \qquad \mathrm{with}~~ m_1 \in [5-150]\msun ~.
\end{equation}
The mass ratio between the secondary and the primary member of the binary $q=m_2/m_1$ is derived as \citep{Sana2012} 
\begin{equation}
	\mathfrak{F}(q)~ \propto ~q^{-0.1} \qquad ~~~\mathrm{with}~~~q = \frac{m_2}{m_1}~ \in [0.1-1]~.
\end{equation}
The orbital period $P$ and the eccentricity $e$ are also extracted according to \cite{Sana2012}:
\begin{equation}
	\mathfrak{F}(\mathscr{P}) ~\propto~ (\mathscr{P})^{-0.55} ~~\mathrm{with}~ \mathscr{P} = \mathrm{log_{10}}(P/\mathrm{day}) \in [0.15-5.5]
\end{equation} 
and 
\begin{equation}
	\mathfrak{F}(e) ~\propto ~e^{-0.42} \qquad ~~\mathrm{with}~~~ 0\leq e < 1.~
\end{equation}

We ran six sets of simulations (see Tab.~\ref{tab:sims}), in order to test the effect of natal kicks and CE efficiency on the formation of double compact-object binaries (DNSs, BHNSs and BHBs). In particular, we assume three different values of the parameter of CE efficiency $\alpha{}=1,\,{}3$ and 5, and we draw iron core-collapse SN kicks from a Maxwellian distribution with 1D rms $\sigma{}_{\rm CCSN}=15$ and 265 km s$^{-1}$.

For each set of simulations we considered 12 sub-sets with different metallicities $Z=0.0002$, $0.0004$, $0.0008$, $0.0012$, $0.0016$, $0.002$, $0.004$, $0.006$, $0.008$, $0.012$, $0.016$ and $0.02$. In each sub-set, we simulated $10^7$ binary systems. Thus, each set of simulations is composed of $1.2\times10^8$ massive binaries.

\section{Results}
\begin{center}
	\begin{figure*}
		\includegraphics[width=17cm]{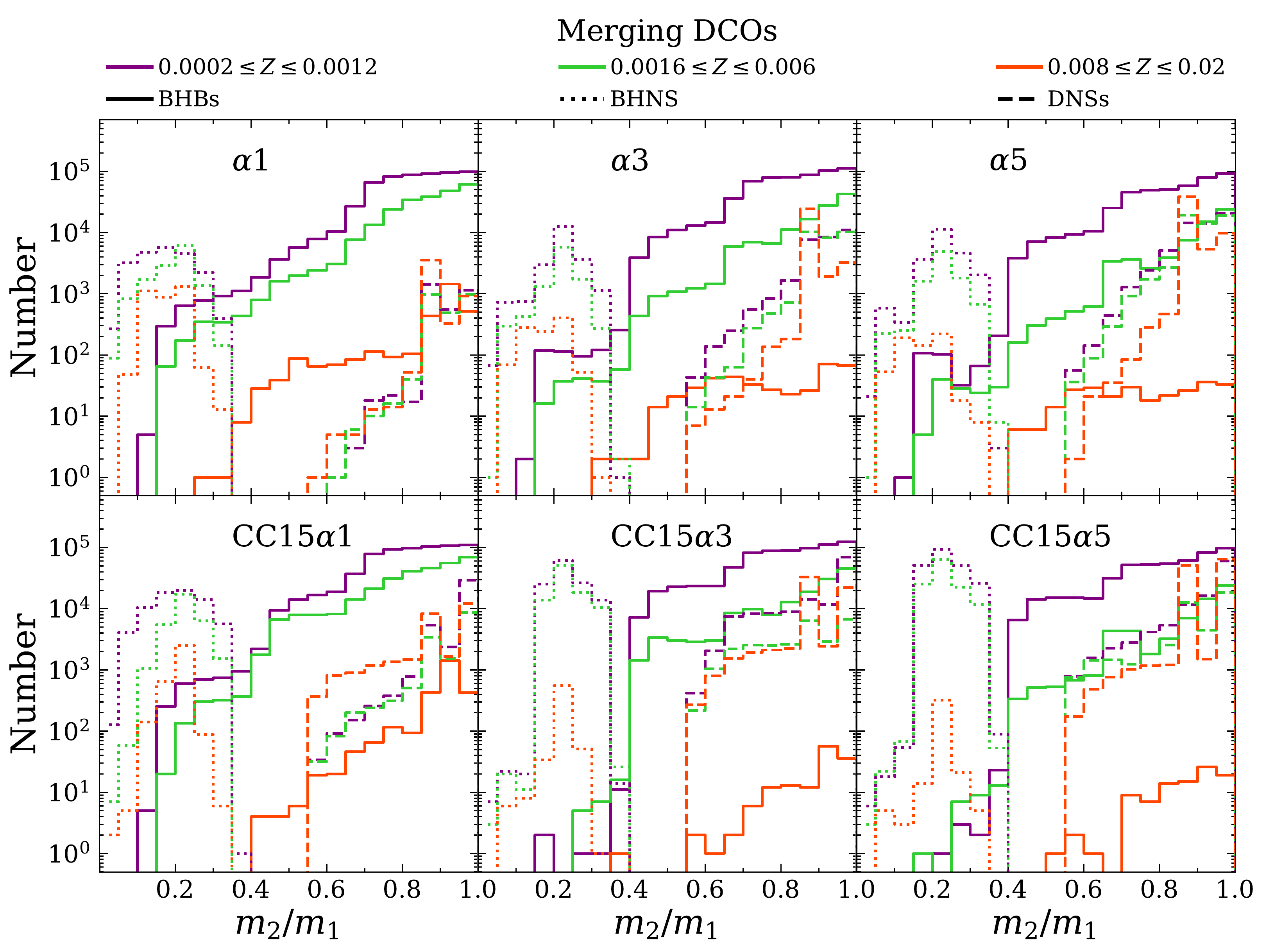}
		\caption{\label{fig:mratio}Mass ratio of the secondary to the primary member of merging double compact objects (DCOs) in all our simulations. Dashed lines: DNSs; dotted lines: BHNSs; solid lines: BHBs. Violet histogram: merging DCOs with metallicity  $0.0002\leq{}Z\leq{}0.0012$; green histogram: merging DCOs with metallicity $0.0016\leq{}Z\leq{}0.006$; red histogram: merging DCOs with metallicity $0.008\leq{}Z\leq{}0.02$. The bin width is $0.05$.}
	\end{figure*}	
\end{center}
\begin{center}
	\begin{figure*}
		\includegraphics[width=13cm]{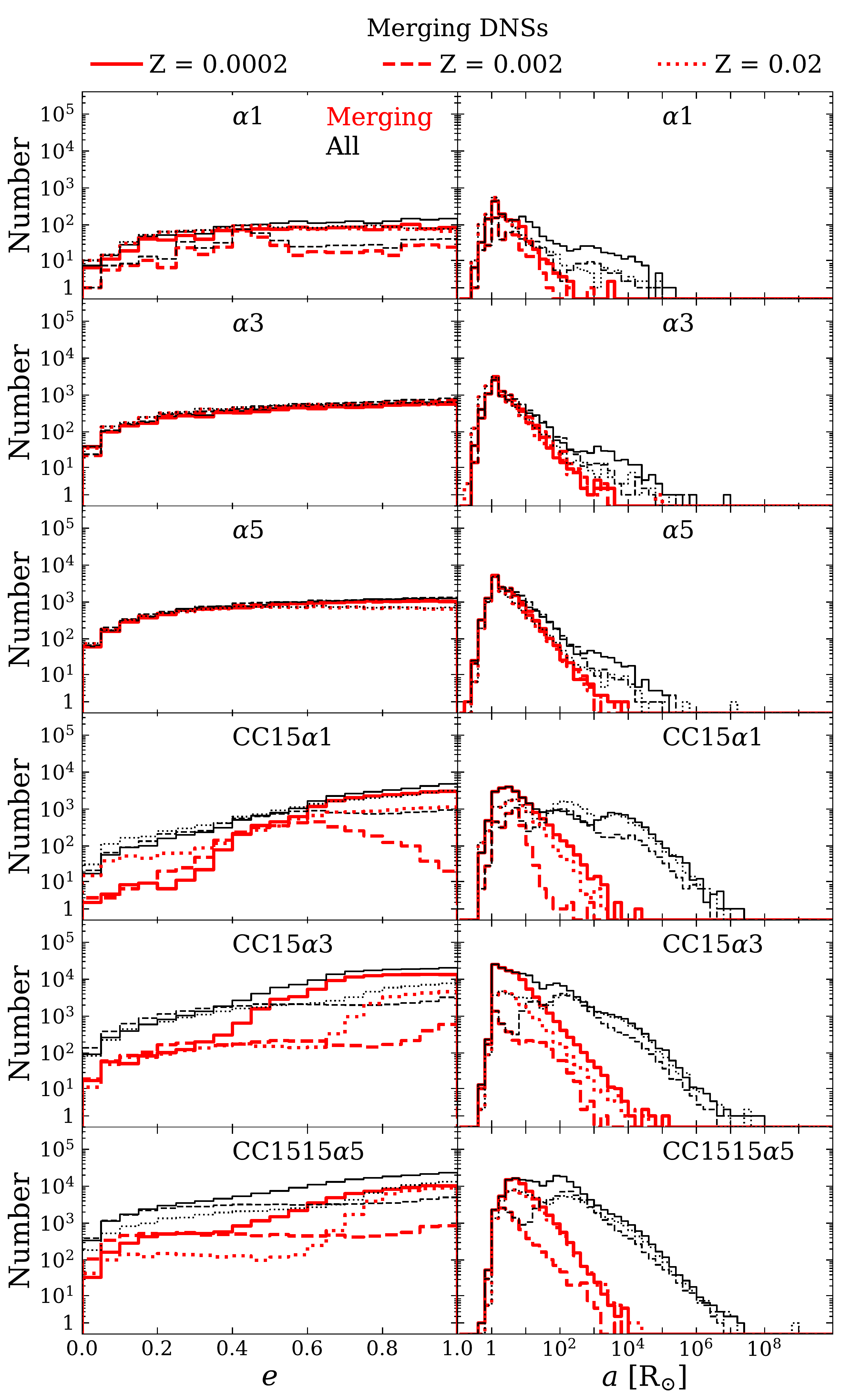}
		\caption{\label{fig:eaDNS}
                  Eccentricity (left-hand panels) and semi-major axis (right-hand panels) of DNSs after the second SN explosion has led to the formation of the second NS (hereafter: initial eccentricity and initial semi-major axis). The bin width is $0.05$ for the eccentricity and $\log(a/{\rm R}_\odot{})=0.2$ for the semi-major axis. All simulations are shown. Black thin lines: all simulated DNSs. Red thick lines: DNSs merging within a Hubble time. Solid lines: $Z=0.0002$; dashed lines: $Z=0.002$; dotted lines $Z=0.02$.}
	\end{figure*}	
\end{center}

\begin{center}
	\begin{figure*}
		\includegraphics[scale=0.5]{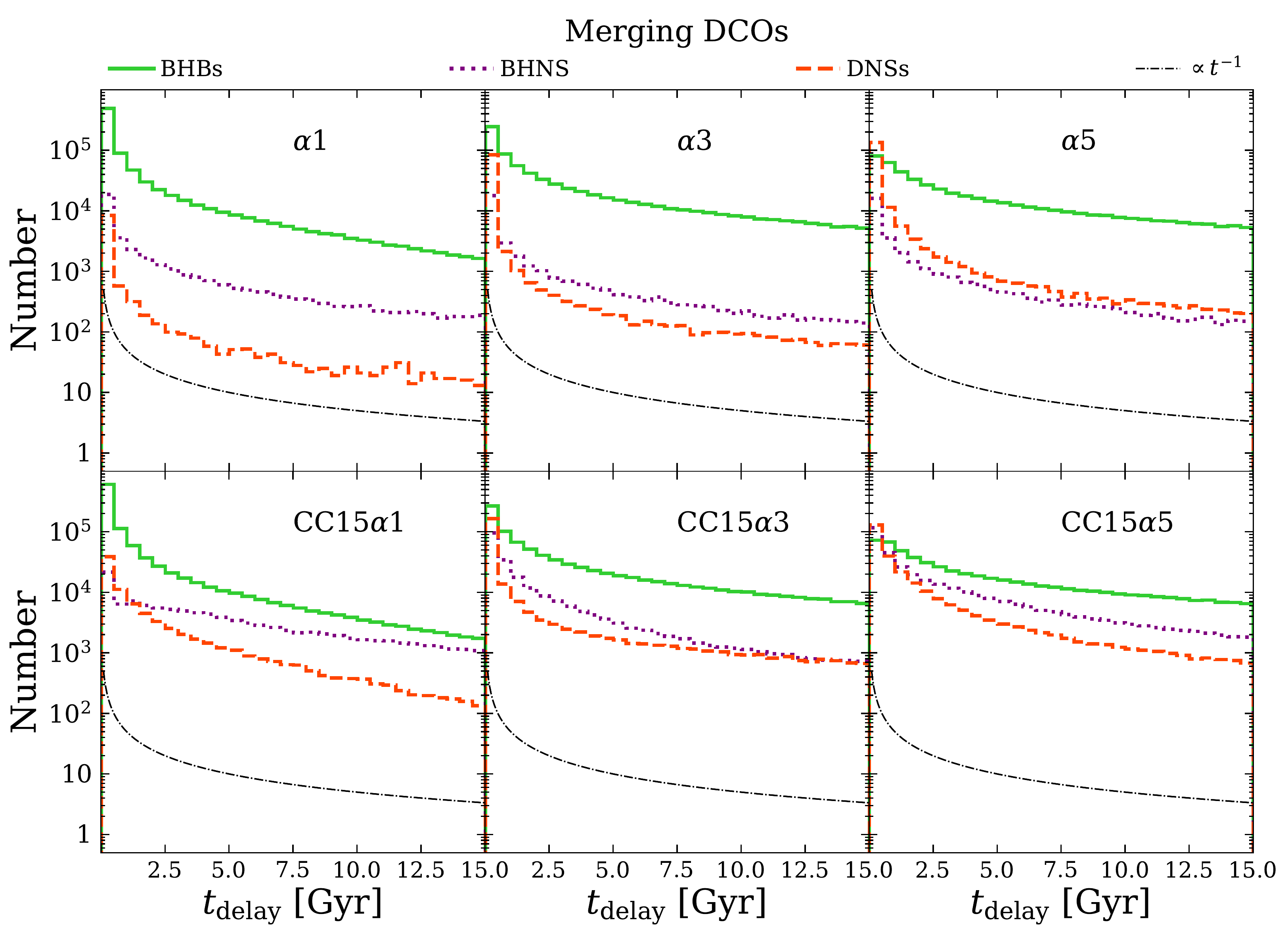}
		\caption{\label{fig:figtdelay} Delay time $t_{\rm delay}$ distribution for DNSs (dashed red line), BHNSs (dotted violet line) and BHBs (green solid line) for all simulation sets. We sum over different metallicities. Thin dot-dashed black line: $dN/dt\propto{}t^{-1}$. The bin width is $0.5$ Gyr.}
	\end{figure*}	
\end{center}

\subsection{Masses and orbital properties of double neutron stars (DNSs)}
Figure~\ref{fig:mtot_DNS} shows the total mass of DNSs ($m_{\rm tot}=m_1+m_2$, where $m_1$ and $m_2$ are the masses of the primary and of the secondary NS, respectively) in simulations $\alpha{}5$ (with large core-collapse SN kicks) and CC15$\alpha{}5$ (with low SN kicks). In particular, the left-hand panels show the mass distribution of all DNSs formed in our simulations $\alpha{}5$ (top) and CC15$\alpha{}5$ (bottom), while the right-hand panels show the mass distribution of DNSs merging within a Hubble time.

\begin{center}
	\begin{figure*}
		\includegraphics[width=17cm]{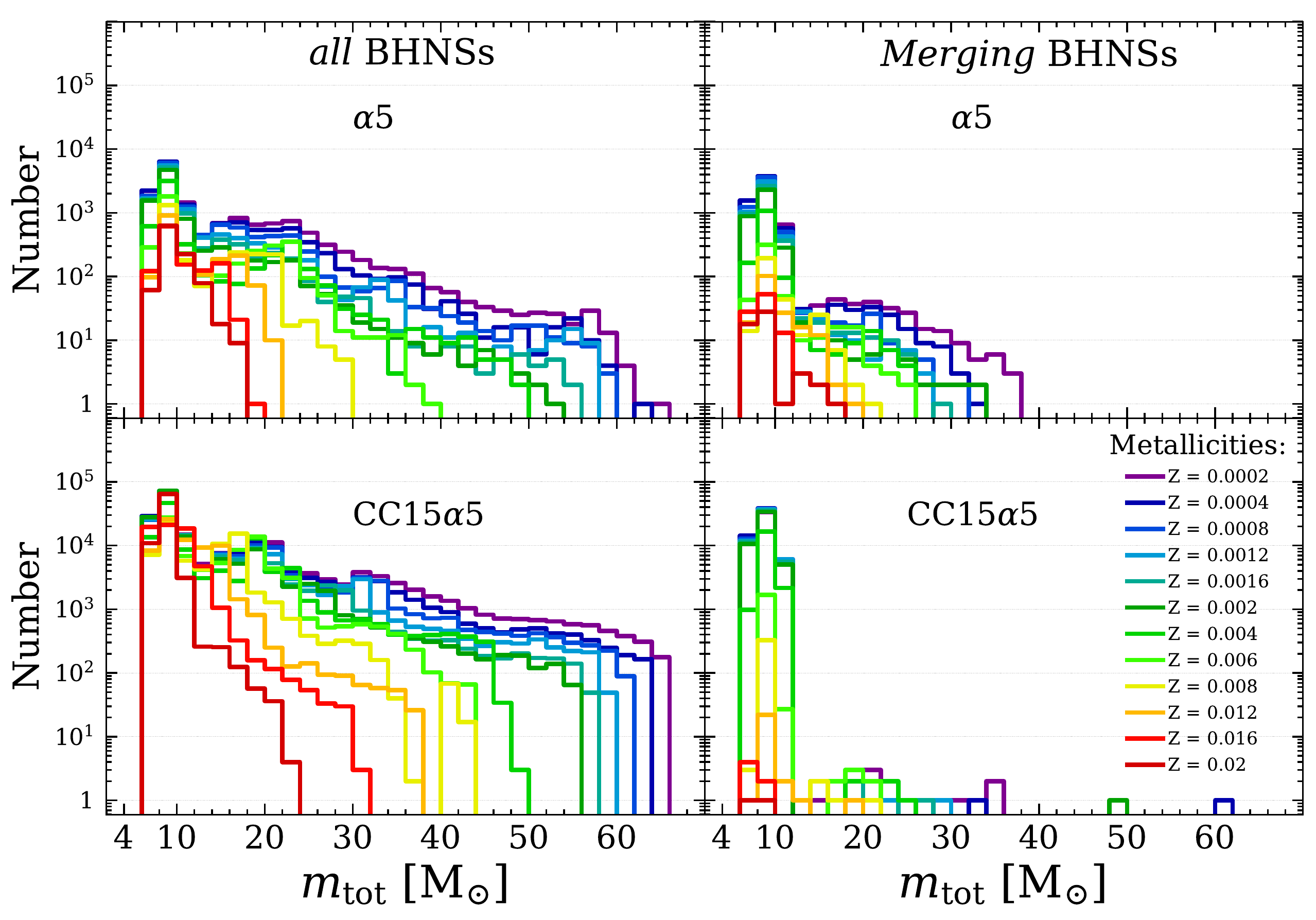}
		\caption{\label{fig:mtot_BHNS}Same as Figure~\ref{fig:mtot_DNS} but for BHNS systems. The bin width is $2$ \msun.}
	\end{figure*}	
\end{center}

\begin{center}
	\begin{figure*}
		\includegraphics[width=14cm]{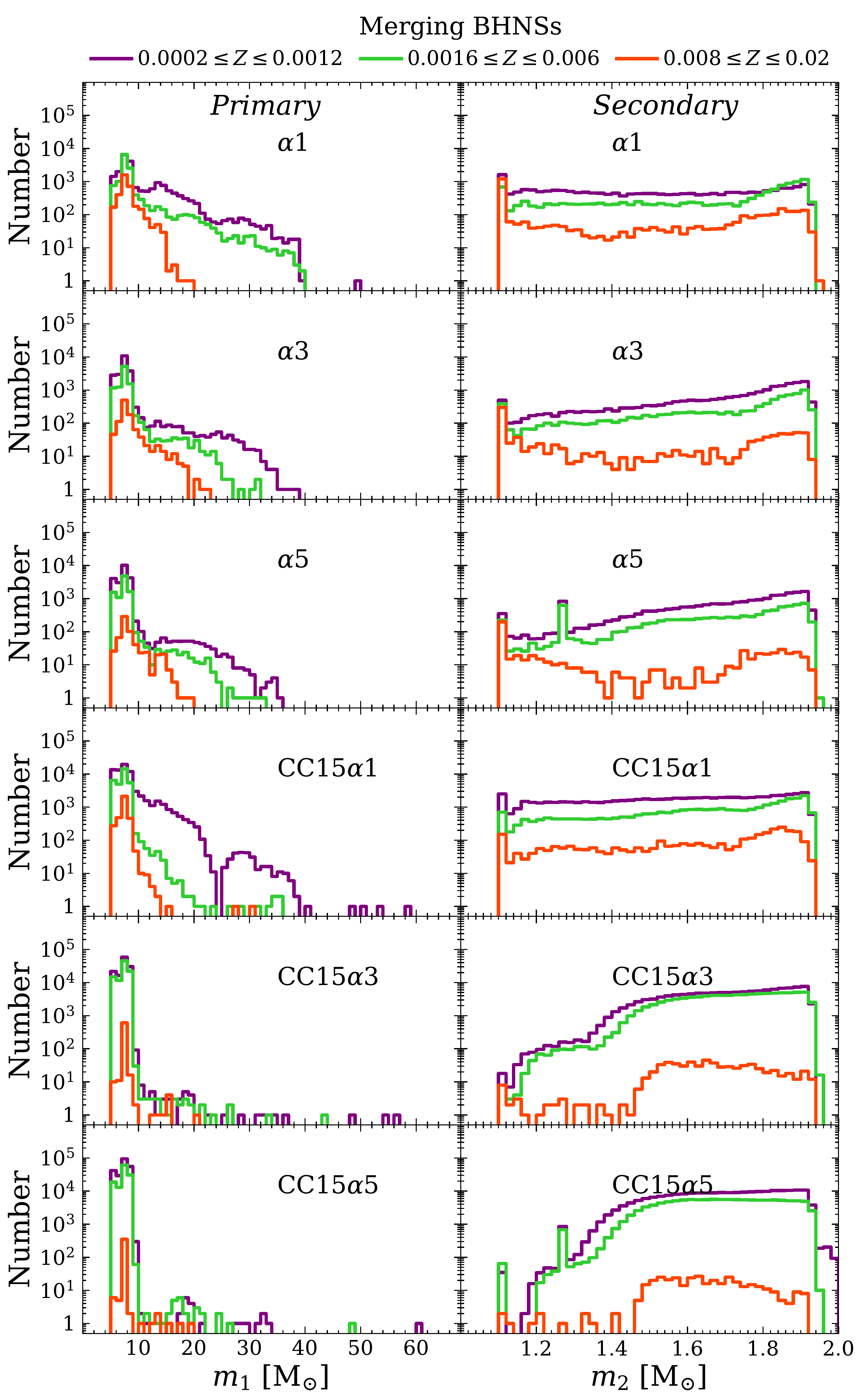}
		\caption{\label{fig:m12BHNS}Same as Figure~\ref{fig:m12DNS} but for BHNS systems. The bin width is $1$ \msun~and $0.02$ \msun~for $m_1$ and $m_2$, respectively.}
	\end{figure*}	
\end{center}

In simulation $\alpha{}5$, most  ($\sim 83-96$ per cent) DNSs merge within a Hubble time. The mass distribution of merging DNSs is not significantly different from the mass distribution of all DNSs. DNSs have a total mass ranging from 2.2 to $\sim{}4$ M$_\odot$, but lower masses ($m_{\rm tot}\lesssim{}2.75$ M$_\odot$) are more common than large masses. The distributions have two peaks at $m_{\rm tot}\sim{}2.2$ and 2.4 M$_\odot$, corresponding to the mergers of two NSs  with mass $m_1\sim{}m_2\sim{}1.1$ M$_\odot$ and $m_1\sim{}1.3$, $m_2\sim{}1.1$ M$_\odot$, respectively. These two favoured masses come from the adopted SN model. In particular, 1.1 M$_\odot$ is the minimum mass of a NS formed from a core-collapse SN according to the rapid SN model of \cite{Fryer2012}, while $\sim{}1.26$ M$_\odot$ is the minimum mass of a NS formed from an electron-capture SN according to \cite{Fryer2012}. Larger masses are indicative of some amount of fallback, while mass accretion from a companion is mostly negligible (in agreement with e.g. \citealt{Tauris2017}). In simulation $\alpha{}5$, the mass distribution of both merging and all DNSs slightly depends on the metallicity of the progenitor star: lower metallicities tend to produce more massive NSs, because fallback is slightly more efficient. 

In simulation CC15$\alpha{}5$ (representative of the simulations assuming a low natal kick for core-collapse SNe), the number of merging DNSs is significantly lower than the number of all DNSs: only $15-50$ per cent of DNSs merge within a Hubble time. This happens because most systems survive the explosion of a core-collapse SN with such a low kick, including DNSs with large orbital separations, while in simulation $\alpha{}5$ only the closest systems (which are more likely to merge within a Hubble time) remain bound after the core-collapse SN explosion.

In simulation CC15$\alpha{}5$ there is also a clear difference between the mass range of all DNSs and the mass range of DNSs merging within a Hubble time: the former span from 2.2 to 4 M$_\odot$, while the latter range from 2.2 to 3.75 M$_\odot$ and very few merging DNSs have $m_{\rm tot}>3.2$ M$_\odot$. Similar to $\alpha{}5$, low mass systems ($m_{\rm tot}\lesssim{}2.75$ M$_\odot$) are more common and there are two peaks corresponding to  $m_{\rm tot}\sim{}2.2$ and 2.4 M$_\odot$. There is only a slight trend with the metallicity of the progenitor stars. 

The total mass of GW170817 is also shown in Figure~\ref{fig:mtot_DNS} \citep{Abbott2017d}. In our simulations we form a large number of systems with mass consistent with GW170817. The number of systems in the mass bin consistent with the total mass of GW170817 is ($\sim{}2-10$) $\sim{}5-20$ per cent the number of (merging) systems in the peak of the distribution.

Figure~\ref{fig:m12DNS} shows the mass of the primary  and the mass of the secondary member of DNSs which merge within a Hubble time, for all simulations considered in this paper. For primary (secondary) we mean the most (least) massive member of the binary. This is independent of which NS forms first.

For each simulation, we gather different metallicities into three groups: $0.0002\leq{}Z\leq{}0.0012$, $0.0016\leq{}Z\leq{}0.006$, $0.008\leq{}Z\leq{}0.02$. In all cases we clearly see the two peaks at $\sim{}1.1$ M$_\odot$ (minimum mass of NSs formed from iron core-collapse SNe in the rapid model) and $\sim{}1.26$ M$_\odot$ (minimum mass of NSs formed from electron-capture SNe) for both the primary and the secondary NS. There is a weak trend with metallicity in simulations with large $\alpha$ ($\alpha{}3$, $\alpha{}5$, CC15$\alpha{}3$ and CC15$\alpha{}5$), while simulations with $\alpha{}=1$ ($\alpha{}1$ and CC15$\alpha{}1$) show a more uncertain behaviour.

Primary masses span from $\sim{}1.1$ to $\sim{}2.0$ M$_\odot$ in all simulations. As for the secondary masses, there is a clear difference between simulations with large core-collapse SN kicks ($\alpha{}1$, $\alpha{}3$ and $\alpha{}5$) and simulations with low core-collapse SN kicks (CC15$\alpha{}1$, CC15$\alpha{}3$ and CC15$\alpha{}5$). In the latter simulations, the secondary NSs have generally lower mass than in the former ones. The main reason is that lower mass NSs in our simulations form mostly from iron core-collapse SNe, while high mass NSs form mainly from electron-capture SNe. If the core-collapse SN kicks are lower, the majority of merging secondary NSs forms through core-collapse SNe.

\begin{center}
	\begin{figure*}
		\includegraphics[scale=0.52]{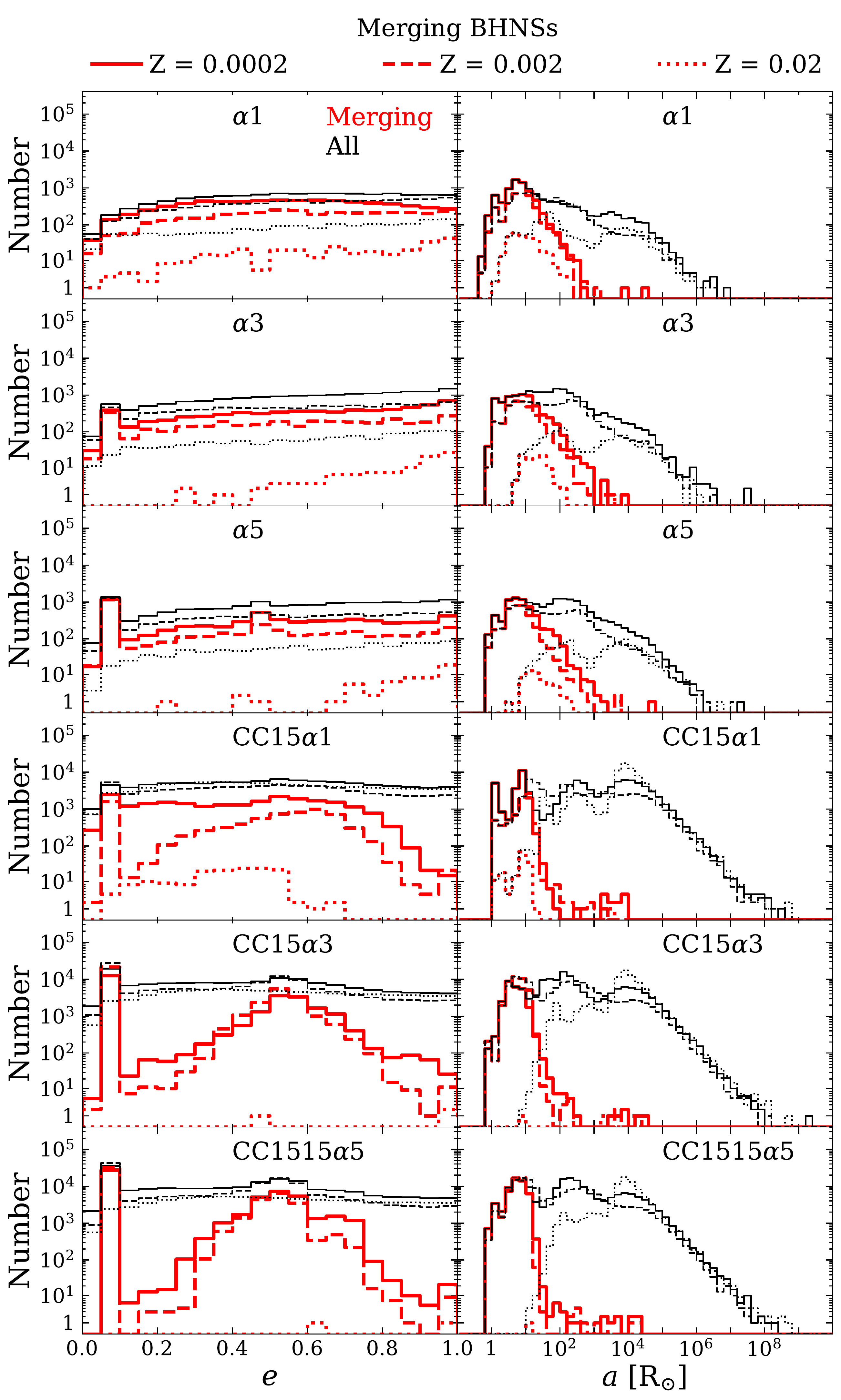}
		\caption{\label{fig:eaBHNS}Same as Figure~\ref{fig:eaDNS} but for BHNS systems.}
	\end{figure*}	
\end{center}

Figure~\ref{fig:mratio} shows the ratio of the secondary to the primary NS mass (dashed lines) for merging DNSs in our simulations. The mass ratio of merging DNSs is always $>0.5$.

Figure~\ref{fig:eaDNS} shows the eccentricity ($e$) and the semi-major axis ($a$) of DNSs after the second SN has taken place, i.e. as soon as the simulated system has become a DNS. While these quantities are hardly comparable to any observations, they are useful, on a theoretical ground, to understand the formation pathways of double compact objects. From Figure~\ref{fig:eaDNS} we see that DNSs form with relatively large initial eccentricities (with respect to BHNSs and BHBs), as an effect of the natal kick. As we could expect, most merging DNSs have small ($<10^3$ R$_\odot$) semi-major axis after the second SN explosion. There is a clear difference between the runs with low SN kicks and those with high SN kicks: in the former a high number of NSs have very large semi-major axes ($10^2-10^8$ R$_\odot$) and do not merge unless they have extreme eccentricity. This explains why much less than 50 per cent of all DNSs merge in the simulations with low core-collapse SN kicks.

\begin{center}
	\begin{figure*}
		\includegraphics[scale=0.5]{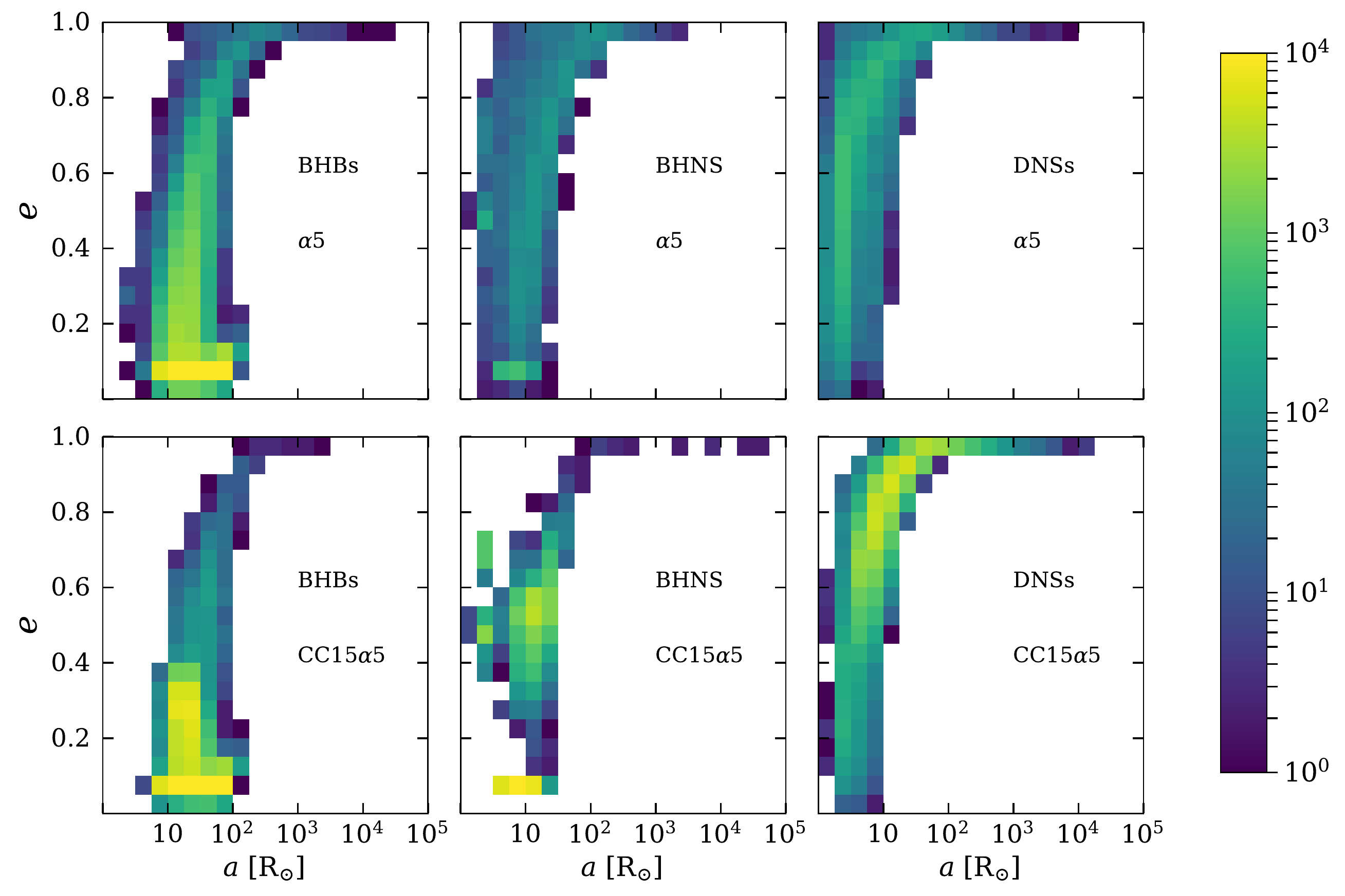}
		\caption{\label{fig:eacontoursBHNS} Initial semi-major axis versus initial eccentricity of merging DCOs (BHBs, BHNSs and DNSs). The logarithmic colour-coded map indicates the number of merging systems. Top (bottom) panels show simulations with $\alpha = 5$ and high-velocity (low-velocity) core-collapse SN kicks. From left to right: BHBs, BHNSs and DNSs. The bin width is $0.05$ for the eccentricity and $\log(a/{\rm R}_\odot{})=0.5$ for the semi-major axis.}
	\end{figure*}	
\end{center}

Finally, Figure~\ref{fig:figtdelay} shows the distribution of the delay time $t_{\rm delay}$ (i.e. the time elapsed from the formation of the progenitor binary to the merger of the two compact objects). Previous studies (\citealt{Belczynski2016,Lamberts2016,Mapelli2017}) indicate that the distribution of delay times should approximately scale as $t^{-1}$. Overall, this trend $dN/dt\propto{}t^{-1}$ is confirmed in our simulations, but we notice also a slight dependence of $t_{\rm delay}$ distribution on $\alpha{}$ and on the natal kicks.

\subsection{Masses and orbital properties of BHNSs}
Figure~\ref{fig:mtot_BHNS} shows the total mass distribution of BHNSs in simulations $\alpha{}5$ and CC15$\alpha{}5$. If we consider all BHNSs formed in our simulations (left-hand panels), the total masses of BHNSs span from $\sim{}6$ to $\sim{}66$ M$_\odot$ in both $\alpha{}5$ and CC15$\alpha{}5$, although the distributions peak at relatively low masses ($\sim{}6-14$ M$_\odot$). There is a clear trend with metallicity for both $\alpha{}5$ and CC15$\alpha{}5$, which is essentially due to the dependence of BH masses on stellar winds (see e.g. \citealt{Giacobbo2018}).

If we consider only those BHNSs which merge within a Hubble time (right-hand panels), we find that low-mass BHNSs are more likely to merge than high-mass BHNSs. In the case $\alpha{}5$, the maximum total mass of merging BHNSs is $\lesssim{}40$ M$_\odot$ and the most common systems have mass $\sim{}6-12$ M$_\odot$. In the case CC15$\alpha{}5$ the situation is even more extreme: almost all merging BHNSs have mass $\sim{}6-12$ M$_\odot$ and only few systems have larger mass (up to $\sim{}60$ M$_\odot$). The behaviour of BHNS total mass distributions in the other simulations is intermediate between that of $\alpha{}5$ and that of CC15$\alpha{}5$.
\begin{center}
	\begin{figure*}
		\includegraphics[scale=0.5]{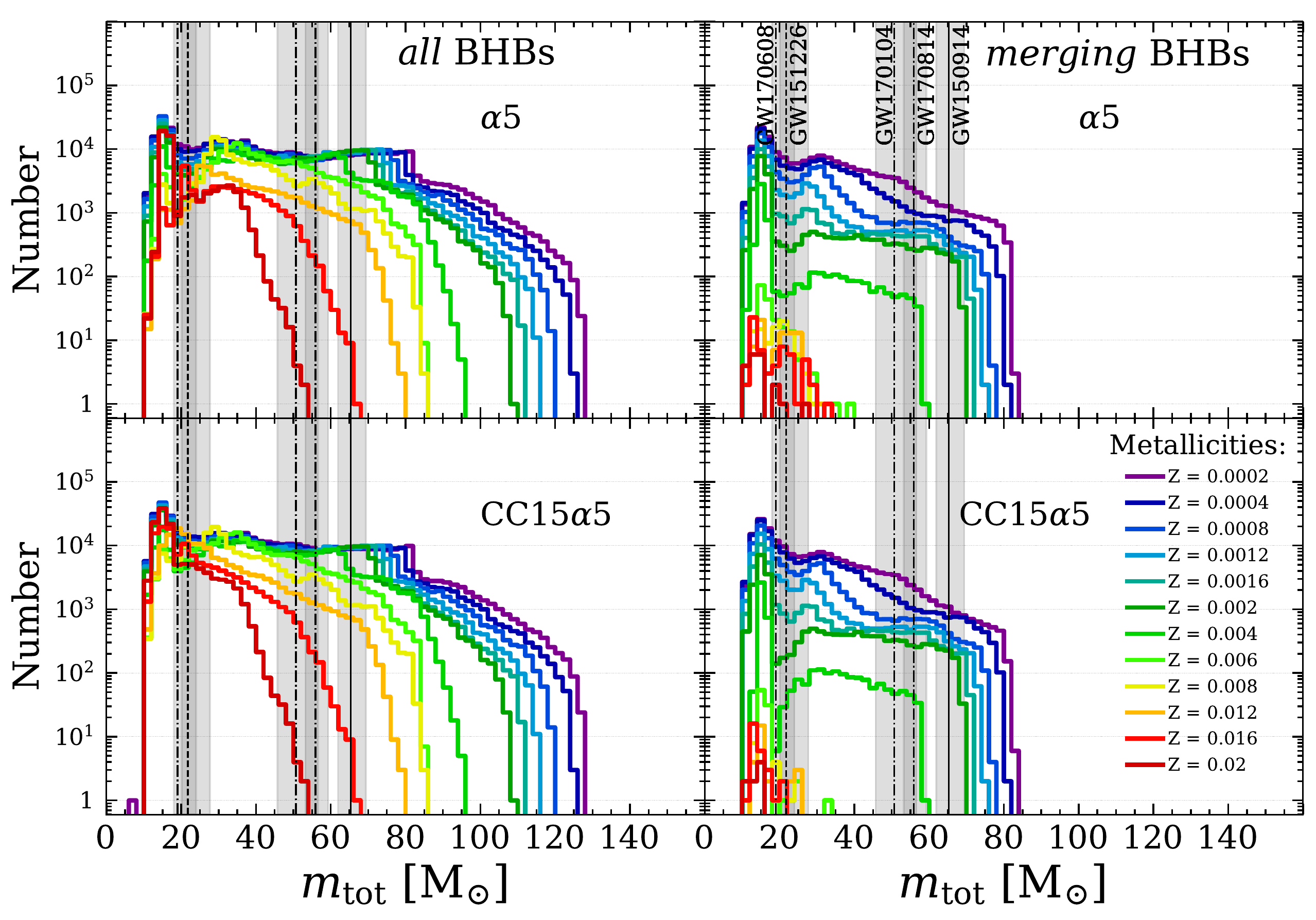}
		\caption{\label{fig:mtot_BHB}Same as Figure~\ref{fig:mtot_DNS} but for BHBs. The vertical lines are the median values of the total masses of the GW events and their 90 per cent credible intervals: GW150914, GW151226~\citep{Abbott2016d}, GW170104~\citep{Abbott2017a}, GW170608~\citep{Abbott2017c} and GW170814~\citep{Abbott2017b}. The bin width is $2$ \msun. Note that the distribution of the observed masses will favor higher value of the mass because of observation selection effects.}
	\end{figure*}	
\end{center}
\begin{center}
	\begin{figure*}
		\includegraphics[scale=0.52]{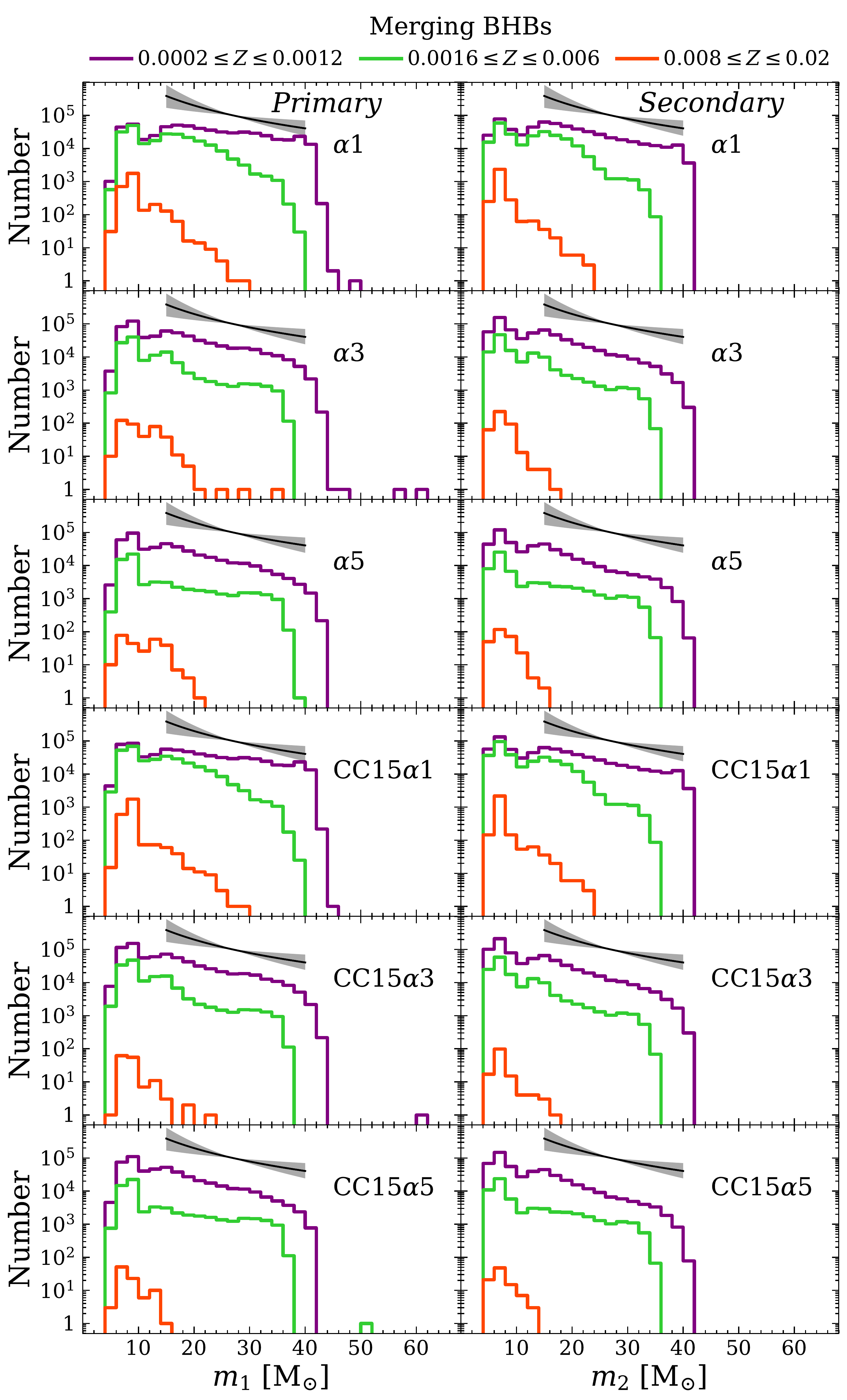}
		\caption{\label{fig:m12BHB}Same as Figure~\ref{fig:m12DNS} but for BHBs. The black line in each panel represents the power-law index inferred from the mass distribution of GW events observed in O1 plus GW170104 ($2.3^{+1.3	}_{-1.4}$, \citealt{Abbott2017a}) and the shadowed region is the corresponding 90 per cent credible interval. The bin width is $2$ \msun{} for both $m_1$ and $m_2$.}
	\end{figure*}	
\end{center}

Figure~\ref{fig:m12BHNS} shows the mass distribution of the BH (left-hand panels) and of the NS (right-hand panels) for BHNSs merging within a Hubble time. From this Figure, we see that the simulations CC15$\alpha{}3$ and CC15$\alpha{}5$ are the most extreme cases: most of the BHs in merging BHNSs have mass $<15$ M$_\odot$ and most of the NSs have mass $\gtrsim{}1.3$ M$_\odot$. 

Even in the other simulations ($\alpha{}1$, $\alpha{}3$, $\alpha{}5$ and CC15$\alpha{}1$), the NS mass distribution in the case of BHNSs is skewed toward significantly higher values than in the case of merging DNSs. In all simulations we see a trend with metallicity: more merging BHNSs form from metal-poor progenitors than from metal-rich ones.

Figure~\ref{fig:mratio} shows that the ratio between the NS mass and the BH mass in merging BHNSs is always $q\lesssim{}0.4$ ($q=0.4$ is the maximum possible value by construction, because the maximum possible mass of a NS is $\sim{}2$ M$_\odot$ and the minimum possible mass of a BH is $\sim{}5$ M$_\odot$, according to the rapid core-collapse SN model by \citealt{Fryer2012}), while the most likely value is $\approx{}0.2$.

Finally, Figure~\ref{fig:eaBHNS} shows the eccentricity and the semi-major axis of the simulated BHNSs just after both compact objects have formed (hereafter: initial eccentricity and initial semi-major axis). The initial semi-major axes of merging BHNSs are $\lesssim{}10^3$ R$_\odot$, as in the case of DNSs. We also note that in the three simulations with low core-collapse SN kicks (CC15$\alpha{}1$, CC15$\alpha{}3$ and CC15$\alpha{}5$) merging BHNSs have significantly smaller initial semi-major axes ($\lesssim{}10^2$ R$_\odot$) than in the other runs. This can be explained if we look also at the distribution of initial eccentricities.

The initial eccentricities of merging BHNSs in the simulations with low core-collapse SN kicks have two main peaks: a sharp peak at $e\sim{}0$ and a broad peak at $e\sim{}0.5$.

The sharp peak with $e=0$ is produced by BHNSs in which the BH forms after the NS and is born by direct collapse. In this case the BH is assigned low or no natal kick \citep{Fryer2012} and the binary, which has circularized during previous evolution, remains circular. 

In contrast, the broad peak at $e\sim{}0.5$ is populated mostly by BHNSs in which the BH forms before the NS. This peak originates from the combination of two effects. First, there are very few BHNSs with semi-major axis $\lesssim{}10^3$ R$_\odot$ and eccentricity $e\gg{}0.5$ (as we can see from Figure~\ref{fig:eacontoursBHNS}), because the SN kick is not strong enough to produce large eccentricities in close BHNSs. If we had a number of such highly eccentric systems with small semi-major axis, they would merge within a Hubble time, but we do not have them because of the low kicks.

Second, we have several systems with semi-major axis $\lesssim{}10^3$ R$_\odot$ and eccentricity $e\ll{}0.5$, but they are expected to merge in a timescale longer than the Hubble time. In fact, we can estimate the merger timescale by GW decay as \citep{Peters1964}
\begin{equation}\label{eq:peters}
  t_{\rm GW}=\frac{5}{256}\,{}\frac{c^5}{G^3}\,{}\frac{a^4\,{}(1-e^2)^{7/2}}{m_1\,{}m_2\,{}(m_1+m_2)},
\end{equation}
where $c$ is the light speed, $G$ the gravity constant, $a$ the binary semi-major axis, $e$ the orbital eccentricity, $m_1$ the mass of the primary and $m_2$ the mass of the secondary. If we substitute  $m_1=7$ M$_\odot$, $m_2=1.5$ M$_\odot$ (which are typical values for our BHNSs, see e.g. Fig.~\ref{fig:m12BHNS}) and $a=10$ R$_\odot$ (which is a very close semi-major axis) into equation~\ref{eq:peters}, we obtain $t_{\rm GW}\sim{}6.2,$ 17 Gyr for $e=0.5$ and 0.0, respectively.

The combination of these two effects produces the broad peak centred at $e\sim{}0.5$.

In simulations $\alpha{}1$, $\alpha{}3$ and $\alpha{}5$ we see no peak at $e\sim{}0.5$, but rather the eccentricities are skewed toward higher values, because the second compact object forms with a higher natal kick, which can unbind the binary or make it very eccentric.

\subsection{Masses and orbital properties of BHBs}
Figure~\ref{fig:mtot_BHB} shows the total mass distribution of BHBs  in simulations $\alpha{}5$ (top) and CC15$\alpha{}5$ (bottom) for all simulated BHBs (left) and only for BHBs merging within a Hubble time (right). This is analogous to Figure~8 of \cite{Giacobbo2018} but here we consider a significantly different model of core-collapse SN, different kick distributions, and a different value of $\alpha{}$.  The main results of \cite{Giacobbo2018} still hold also for this model: if we consider all BHBs formed in our simulations (left-hand panels of Figure~\ref{fig:mtot_BHB}), the maximum total mass\footnote{There is a typo in the $x-$axis label of Figure~8 of \cite{Giacobbo2018}. The most massive BHBs in this figure have $m_{\rm tot}\sim{}130$ M$_\odot$, although from the labels of this Figure it may seem that they have $m_{\rm tot}\sim{}150$ M$_\odot$.} is $\sim{}130$ M$_\odot$.

In contrast, if we look at the total mass of BHBs merging within a Hubble time (right-hand panels of Figure~\ref{fig:mtot_BHB}), we find that the most massive merging BHBs have $m_{\rm tot}\sim{}80-90$ M$_\odot$. The reason why our most massive BHB systems do not merge in a Hubble time is that they come from the evolution of stars which develop large radii during their super-giant phase: if the initial orbital separation of these massive progenitors is large ($\gtrsim{}10^3$ R$_\odot$), they evolve into very massive BHBs which do not merge in a Hubble time. In contrast, if their initial orbital separation is relatively small, the massive progenitors either merge before becoming BHs, or undergo non-conservative mass transfer episodes leading to strong mass loss and to the formation of two smaller BHs. On the other hand, dynamical evolution in star clusters might accelerate the merger of the most massive BHBs which form in our simulations, by means of three-body encounters and exchanges \citep{Mapelli2016}.


As in \cite{Giacobbo2018}, we confirm that there is a strong dependence on metallicity. Not only does the maximum mass of BHBs strongly depend on the progenitor's metallicity, but also the number of BHBs, especially if we look at systems merging in less than a Hubble time. This strong dependence is an effect of stellar evolution: (i) metal-poor stars tend to produce more massive BHs because they lose less mass by stellar winds \citep{Mapelli2009,Mapelli2010,Belczynski2010,Mapelli2013,Spera2015}, while (ii) metal-rich stars tend to develop larger radii than metal-poor ones and thus are more likely to merge before becoming BHBs; hence, the number of BHBs formed from metal-poor progenitors is larger than that of BHBs formed from metal-rich ones.

We find no significant differences between the simulation $\alpha{}5$  and CC15$\alpha{}5$, because most BHBs get a low natal kick in all our models, given our assumption that the kick is modulated by the amount of fallback (see e.g. \citealt{Giacobbo2018}).

Figure~\ref{fig:m12BHB} shows the mass of the primary (i.e. the more massive) and the secondary (i.e. the less massive) BH in BHBs merging within a Hubble time for all simulations discussed in this paper. In all cases, we see a clear trend with metallicity, in terms of both maximum BH mass and number of merging BHBs.
Especially at low metallicity ($0.0002\leq{}Z\leq{}0.0012$), the mass distribution is broadly consistent with a power law with slope $\approx{2.3}$, in good agreement with the one inferred from GW events \citep{Abbott2016d}.


Figure~\ref{fig:mratio} confirms that merging BHBs in our simulations have preferentially a large mass ratio, but systems with any possible mass ratio down to $m_2/m_1\sim{}0.1$ can form, consistent with \cite{Giacobbo2018}.

\begin{center}
	\begin{figure*}
		\includegraphics[scale=0.52]{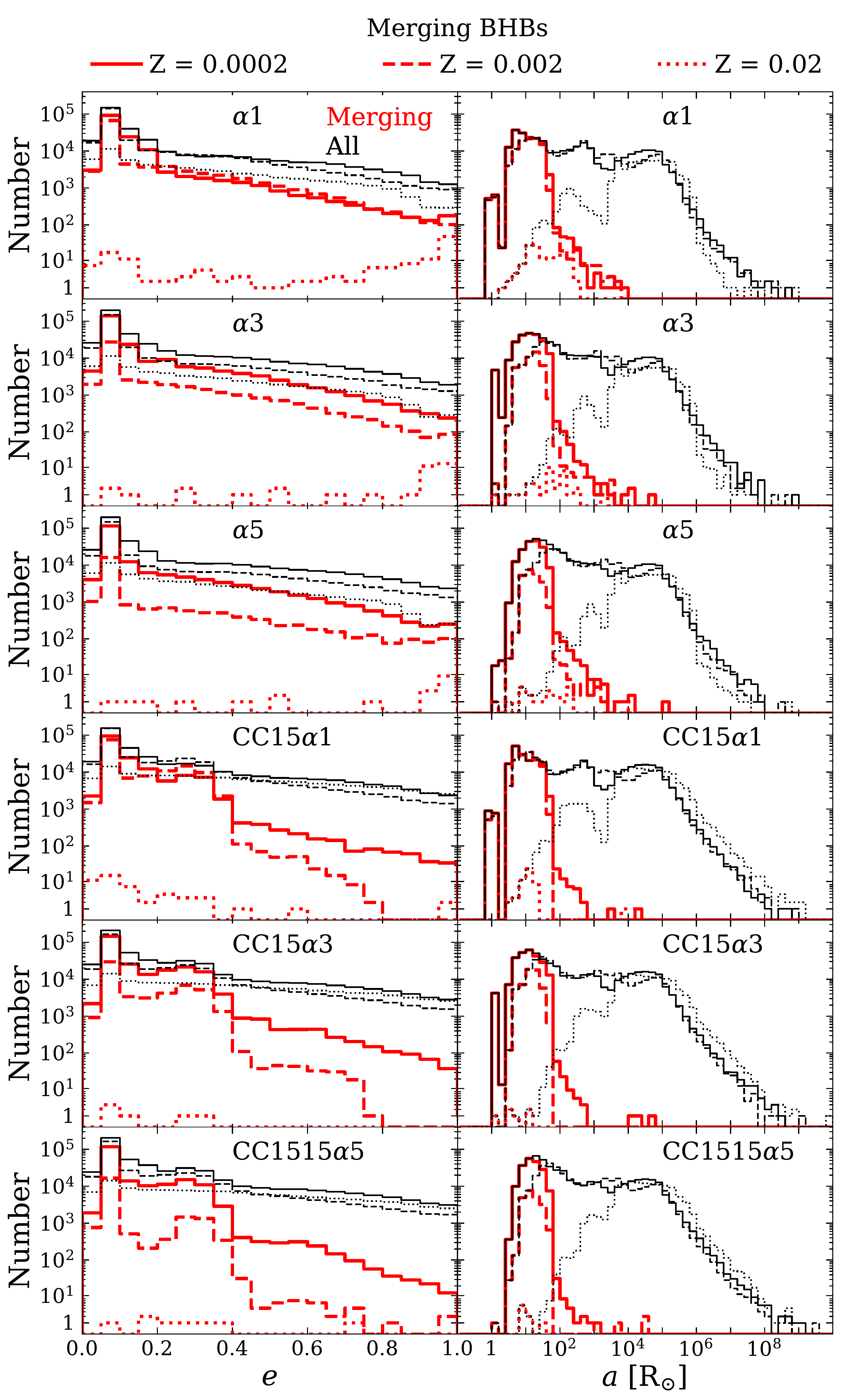}
		\caption{\label{fig:eaBHB} Same as Figure~\ref{fig:eaDNS} but for BHBs.}
	\end{figure*}	
\end{center}

Finally, Figure~\ref{fig:eaBHB} shows the initial eccentricity distribution of BHBs. Most BHBs have relatively low eccentricity, because SN kicks are assumed to be low, depending on the fallback. The initial semi-major axis of merging BHs is generally $\lesssim{}10^2$ R$_\odot$, with a tail of eccentric systems merging with initial semi-major axis up to $\sim{}10^4$ R$_\odot$.

\begin{center}
	\begin{figure*}
		\includegraphics[scale=0.52]{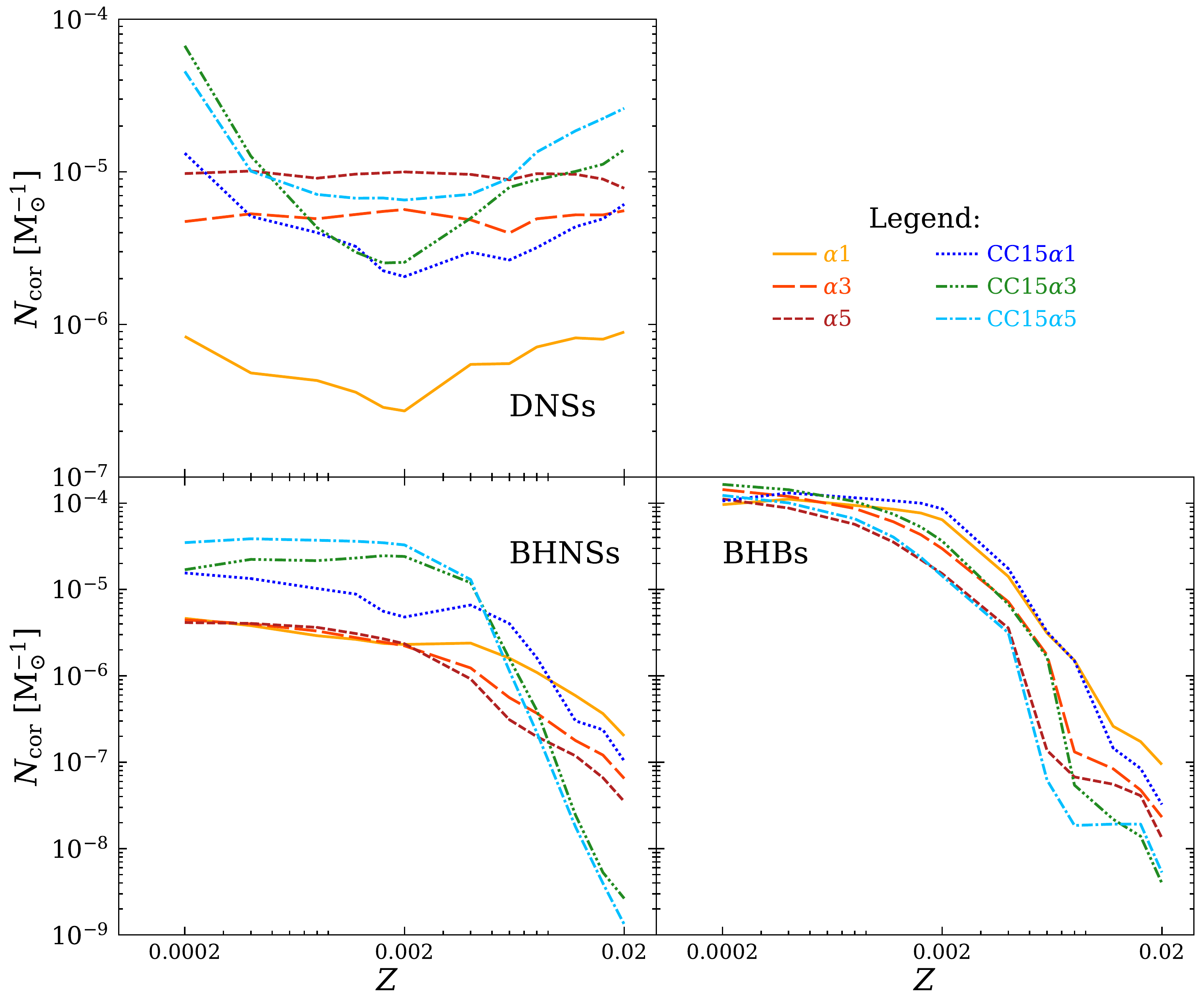}
		\caption{\label{fig:ncor}Number of merging DCOs per unit mass ($N_{\rm cor}$) as a function  of progenitor's metallicity for all our simulations (see Table~\ref{tab:sims}). Top-left panel: merging DNSs; bottom-left panel: BHNS; bottom-right panel: BHBs.}
	\end{figure*}	
\end{center}

Another interesting feature of merging BHBs is the dependence of $t_{\rm delay}$ on the value of $\alpha{}$ (Figure~\ref{fig:figtdelay}): the number of systems with short delay time increases significantly for a smaller value of $\alpha{}$. Small values of $\alpha{}$ correspond to a more efficient shrinking of the binary system during CE, resulting in a faster merger. This effect appears to be particularly important for BHBs.

\subsection{Mergers per unit stellar mass} 
\label{sec:4.3}

For each set of simulations, we compute the number of merging double compact objects (DCOs) per unit mass $N_{\rm cor}$ in the same way as described in our previous work \citep[see][]{Mapelli2017,Giacobbo2018}:
\begin{equation}
	N_{\rm{cor,i}} = \mathrm{f_{bin}\,{}f_{IMF}}\,{} \frac{N_{\mathrm{merger,i}}}{M_{\mathrm{tot,sim}}} \qquad i \in [{\rm DNS, BHNS, BHB}],
\end{equation} 
where $N_{\mathrm{merger,i}}$ is the number of merging DCOs (DNS, BHNS or BHBs); $M_{\mathrm{tot,sim}}$ is the total initial mass of the simulated stellar population; $\rm{f_{bin}}$ is the correction factor used to take into account the fact that we only simulate binary systems (we assume that $50$ per cent of the stars are in binaries $f_{\rm bin}=0.5$ \citealt{Sana2013}); $\rm{f_{IMF}}=0.285$ corrects for the fact that we have simulated only systems which have the primary component more massive than $5 \msun$.

Figure~\ref{fig:ncor} shows $N_{\rm{cor}}$ as a function of the metallicity $Z$ for all the simulations and separately for each type of DCO: DNSs, BHNSs and BHBs. In \citet{Giacobbo2018} we have already shown a similar figure for BHBs, 
but here we consider different SN prescriptions, CE parameters and SN kicks.
          
The number of mergers per unit mass spans a large range of values for both DNSs, BHNSs and BHBs, depending on natal kicks, on CE efficiency and on progenitor's metallicity.

$N_{\rm{cor}}$ strongly depends on progenitor's metallicity for both BHBs and BHNSs. 
In particular, $N_{\rm{cor,BHB}}$ ($N_{\rm cor,BHNS}$) is $\gtrsim 3$ ($\gtrsim{}2$) orders of magnitude higher at low metallicity than at high metallicity. 

In contrast, the number of DNS mergers per unit mass $N_{\rm{cor,DNS}}$ is almost insensitive to progenitor's metallicity in runs~$\alpha{}3$ and $\alpha{}5$. In the other simulations ($\alpha{}1$, CC15$\alpha{}1$, CC15$\alpha{}3$ and CC15$\alpha{}5$) $N_{\rm{cor,DNS}}$ is significantly lower for progenitor metallicity $Z\sim{}0.002$ than for progenitors with higher or lower metallicity. 
This happens because stellar radii in the HG and giant phase tend to reach larger values at $Z\sim{}0.002$ than at other metallicities in {\sc MOBSE}. When SN kicks are low or inspiral by CE is particularly efficient (i.e. $\alpha{}\leq{}1$), a larger number of binaries merge before forming a DNS if the stellar radii are larger. For the same reason, $N_{\rm cor, DNS}$ is larger if $\alpha{}$ is larger (i.e. if CE is less efficient in merging binaries prematurely).

\begin{center}
	\begin{figure*}
		\includegraphics[scale=0.52]{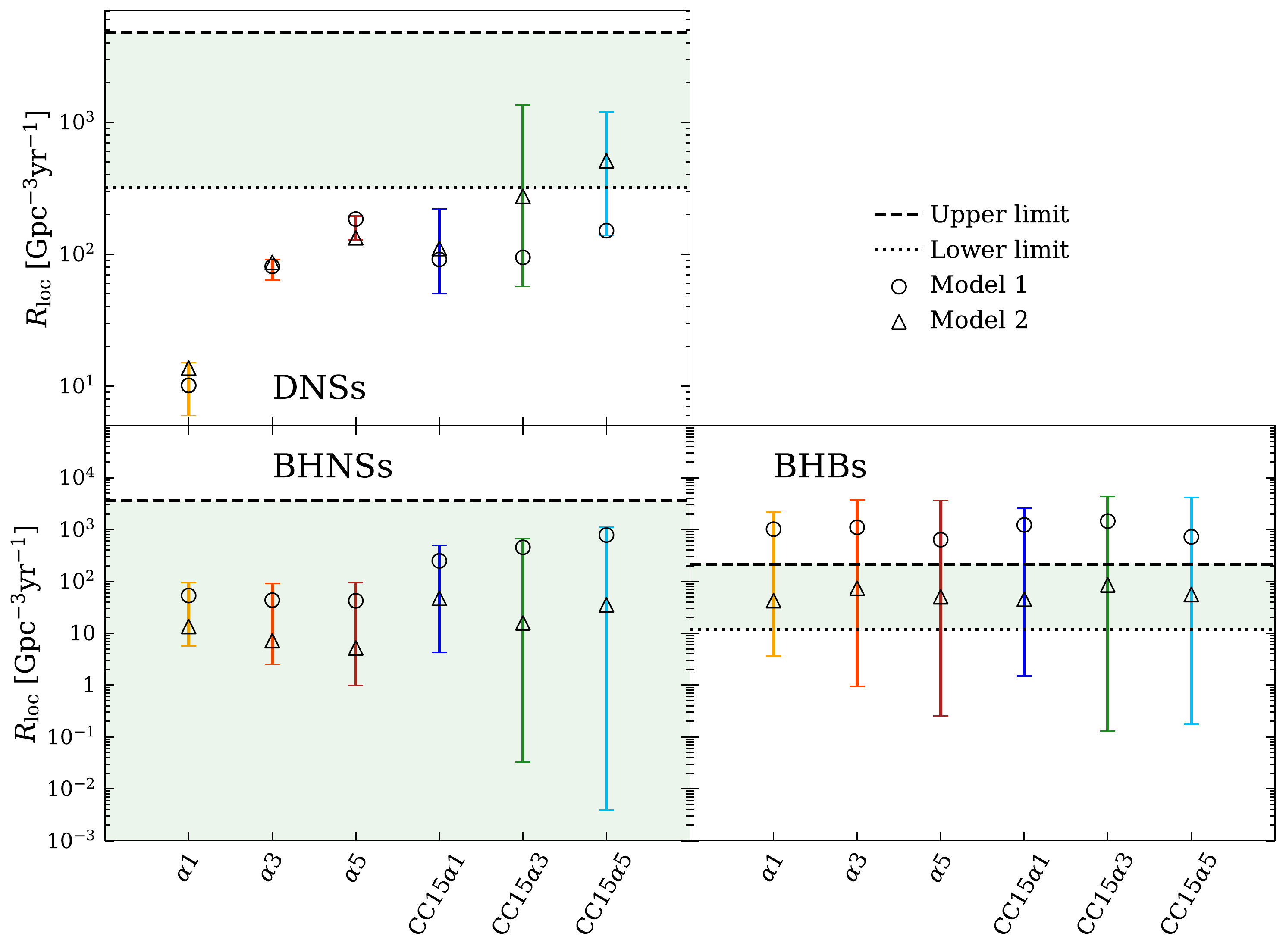}
		\caption{\label{fig:rate} Local merger rate density $R_{\rm loc}$ calculated from equation~\ref{eq:rate} for each simulation.  Top-left: local merger rate density for DNSs. Bottom-left: local merger rate density for BHNSs. Bottom-right: local merger rate density for BHBs.  The error bars show the maximum difference between estimates of $R_{\rm loc}$ for different metallicities. Open circles (open triangles) are $R_{\rm loc}$ obtained assuming $model~1$ ($model~2$) for the cosmic evolution of metallicity. Green shaded areas: local merger rate density inferred from LIGO-Virgo observations (from \citealt{Abbott2017d}, \citealt{Abbott2016e} and \citealt{Abbott2017a} for DNSs, BHNSs and BHBs, respectively).}
	\end{figure*}	
\end{center}

\begin{table}
	\begin{center}
		\caption{Local merger rate density.\label{tab:mergerrate}}
		\begin{tabular}{ccccc}
			\toprule
			{\bf ID} & {\bf Method} & $R_{\rm loc,DNS}$ & $R_{\rm loc,BHNS}$ & $R_{\rm loc,BHB}$ \\
			 &  & \scriptsize{(Gpc$^{-3}$ yr$^{-1}$)} & \scriptsize{(Gpc$^{-3}$ yr$^{-1}$)} & \scriptsize{(Gpc$^{-3}$ yr$^{-1}$)} \\
			\midrule
			  		   & $Z = 0.02$ & $1.3\times 10^{1}$ & $5.7\times 10^{0}$ & $3.6\times 10^{0}$ \\
			  		   & $Z = 0.002$ & $6.0\times 10^{0}$ & $5.5\times 10^{1}$ & $1.8\times 10^{3}$ \\
			  		   $\alpha1$ & $Z = 0.0002$ & $1.5\times 10^{1}$ & $9.5\times 10^{1}$ & $1.4\times 10^{3}$ \\
			  		   & $Model~ 1$ & $1.0\times 10^{1}$ & $5.3\times 10^{1}$ & $1.0\times 10^{3}$ \\
			  		   & $Model~ 2$ & $1.4\times 10^{1}$ & $1.3\times 10^{1}$ & $4.3\times 10^{1}$ \\
			  		   \hline
			  		   & $Z = 0.02$ & $8.6\times 10^{1}$ & $2.5\times 10^{0}$ & $9.4\times 10^{-1}$ \\
			  		   & $Z = 0.002$ & $9.1\times 10^{1}$ & $5.5\times 10^{1}$ & $1.2\times 10^{3}$ \\
			  		   $\alpha3$ & $Z = 0.0002$ & $7.7\times 10^{1}$ & $9.1\times 10^{1}$ & $3.7\times 10^{3}$ \\
			  		   & $Model~ 1$ & $8.1\times 10^{1}$ & $4.3\times 10^{1}$ & $1.1\times 10^{3}$ \\
			  		   & $Model~ 2$ & $8.7\times 10^{1}$ & $7.0\times 10^{0}$ & $7.4\times 10^{1}$ \\
			  		   \hline
			  		   & $Z = 0.02$ & $1.3\times 10^{2}$ & $9.9\times 10^{-1}$ & $2.6\times 10^{-1}$ \\
			  		   & $Z = 0.002$ & $1.9\times 10^{2}$ & $5.7\times 10^{1}$ & $6.1\times 10^{2}$ \\
			  		   $\alpha5$ & $Z = 0.0002$ & $1.8\times 10^{2}$ & $9.5\times 10^{1}$ & $3.6\times 10^{3}$ \\
			  		   & $Model~ 1$ & $1.8\times 10^{2}$ & $4.2\times 10^{1}$ & $6.4\times 10^{2}$ \\
			  		   & $Model~ 2$ & $1.3\times 10^{2}$ & $5.0\times 10^{0}$ & $5.0\times 10^{1}$ \\
			  		   \hline
			  		   & $Z = 0.02$ & $1.4\times 10^{2}$ & $4.2\times 10^{0}$ & $1.5\times 10^{0}$ \\
			  		   & $Z = 0.002$ & $7.8\times 10^{1}$ & $2.0\times 10^{2}$ & $2.3\times 10^{3}$ \\
			  		   CC15$\alpha1$ & $Z = 0.0002$ & $2.2\times 10^{2}$ & $3.7\times 10^{2}$ & $1.5\times 10^{3}$ \\
			  		   & $Model~ 1$ & $9.1\times 10^{1}$ & $2.5\times 10^{2}$ & $1.2\times 10^{3}$ \\
			  		   & $Model~ 2$ & $1.1\times 10^{2}$ & $4.7\times 10^{1}$ & $4.5\times 10^{1}$ \\
			  		   \hline
			  		   & $Z = 0.02$ & $2.4\times 10^{2}$ & $3.3\times 10^{-2}$ & $1.3\times 10^{-1}$ \\
			  		   & $Z = 0.002$ & $5.9\times 10^{1}$ & $6.7\times 10^{2}$ & $1.7\times 10^{3}$ \\
			  		   CC15$\alpha3$ & $Z = 0.0002$ & $1.3\times 10^{3}$ & $3.3\times 10^{2}$ & $4.3\times 10^{3}$ \\
			  		   & $Model~ 1$ & $9.4\times 10^{1}$ & $4.5\times 10^{2}$ & $1.5\times 10^{3}$ \\
			  		   & $Model~ 2$ & $2.7\times 10^{2}$ & $1.6\times 10^{1}$ & $8.6\times 10^{1}$ \\
			  		   \hline
			  		   & $Z = 0.02$ & $5.5\times 10^{2}$ & $3.9\times 10^{-3}$ & $1.8\times 10^{-1}$ \\
			  		   & $Z = 0.002$ & $1.4\times 10^{2}$ & $1.0\times 10^{3}$ & $6.0\times 10^{2}$ \\
			  		   CC15$\alpha5$ & $Z = 0.0002$ & $1.2\times 10^{3}$ & $8.6\times 10^{2}$ & $4.1\times 10^{3}$ \\
			  		   & $Model~ 1$ & $1.5\times 10^{2}$ & $7.8\times 10^{2}$ & $7.2\times 10^{2}$ \\
			  		   & $Model~ 2$ & $5.1\times 10^{2}$ & $3.5\times 10^{1}$ & $5.6\times 10^{1}$ \\
			  		   \bottomrule	
			\bottomrule	
		\end{tabular}
	\end{center}
	{\small Column 1: simulation's name; column 2: method (single metallicity, $Z$, or evolution of metallicity: $model~1$ and $model~2$); columns 3, 4 and 5: local merger rate density of DNSs, BHNSs and BHBs.}
\end{table}
\subsection{Merger rate density}

We developed a simple procedure to estimate the merger rate per unit volume and unit time in the local Universe, based on the number of mergers per unit mass. 

First, we consider the cosmic star formation rate (SFR) density adopting the fit by \citet{Madau2014},
\begin{equation}\label{eq:SFR}
	{\rm SFR}(z) = 0.015 \frac{(1+z)^{2.7}}{1 + [(1+z)/2.9]^{5.6}} ~~\msun {\rm Mpc}^{-3} {\rm yr}^{-1}. 
\end{equation}
We use equation~\ref{eq:SFR} to compute the SFR at different redshifts, from $z=15$  to $z=0.0$ with a step $\Delta z = 0.1$.

Then, we calculate the total mass of stars formed within a specific redshift bin, assuming that the SFR is constant in each bin of redshift. The total mass of stars formed in the bin is multiplied by $N_{\rm cor,BHB}$, $N_{\rm cor,BHNS}$ and $N_{\rm cor,DNS}$ in order to obtain the total number of merging BHBs, BHNSs and DNSs  formed in each redshift bin per unit volume. We repeat these operations for all bins of redshift and for each bin we take only the fraction of merging systems which have a delay time (Figure~\ref{fig:figtdelay}) such that they merge in the local Universe (namely, $z \in [0,0.1]$). By summing them up, we obtain the total number of mergers occurring at $z \leq{}0.1$ per unit volume\footnote{In this work, we estimate the local merger rate within $z\leq{}0.1$. This redshift interval is actually smaller than the current instrumental horizon of LIGO and Virgo for BHBs, which depends on BHB mass. On the other hand, the merger rate density in the comoving frame is expected to grow mildly with redshift, as shown by \cite{Mapelli2017}. We decided not to account for this effect in the current paper, because the uncertainty is dominated by other approximations in the calculation described here (e.g. the assumption about the metallicity).  A more accurate study of the merger rate as a function of redshift is provided in the companion paper \cite{Mapelli2018}.}. Finally, we divide the result by the look-back time corresponding to $z=0.1$. The following formula summarizes the aforementioned procedure: 
\begin{equation}\label{eq:rate}
	R_{\rm loc,i} = \frac{1}{t_{\rm lb}(z=0.1)} \sum _{z=15} ^{0.1} f_{\rm loc}(z)~ {\rm SFR}(z)~ [t_{\rm lb}(z+\Delta z) - t_{\rm lb}(z)]~,
\end{equation}
where $R_{\rm loc,i}$ is the local merger rate per unit volume and unit time, $i$ can be either BHB, BHNS or DNS, $t_{\rm lb}(z)$ is the look-back time  and $f_{\rm loc}(z)$ is the fraction of  merging systems formed at a given redshift $z$ which merge in the local Universe ($z\leq{}0.1$) per unit mass. We calculate $f_{\rm loc}(z)=N_{\rm cor,i}(z_{\rm form}=z,z_{\rm merg}\leq{}0.1)$, where $N_{\rm cor,i}(z_{\rm form}=z,z_{\rm merg}\leq{}0.1)$ is the number of compact objects per unit mass which form at redshift $z_{\rm form}=z$ and merge at redshift $z_{\rm merg}\leq{}0.1$. For our calculations we assume the cosmological parameters from {\it Planck} 2015 \citep{Planck2016}.

In equation~\ref{eq:rate}, the dependence of the merger rate on metallicity is contained in the term $f_{\rm loc}(z)$. Our knowledge of the stellar metallicity evolution in the Universe is quite more uncertain than our knowledge of the SFR density evolution (see e.g. \citealt{Maiolino2008,Mannucci2009,Madau2014} and references therein). Thus, we decided to make minimal assumptions on the metallicity evolution.

In the simplest case, we assume that all stars in the Universe form with the same metallicity $Z$ and we calculate equation~\ref{eq:rate} for each of the 12 metallicities considered in our simulations. Hence, we estimate that the uncertainty of the local merger rate density is the maximum difference between values of $R_{\rm loc,i}$ calculated with all considered metallicities (error bars in Figure~\ref{fig:rate}). This procedure results in relatively small error bars for the DNSs, which do not depend much on progenitor's metallicity (see Figure~\ref{fig:ncor}), while the uncertainty on the merger rate density of both BHBs and BHNSs spans several orders of magnitude.

Then, we adopt two different models for the evolution of metallicity. In {\emph{model~1}}, we assume that all stars formed in a given redshift bin $\Delta{}z$ have the same metallicity and we calculate this metallicity as $\log{Z(z)/Z_{\odot{}}}=-0.19\,{}z-0.74$ if $z \leq 1.5$ and $\log{Z(z)/Z_{\odot{}}}=-0.22\,{}z-0.66$ if $z > 1.5$. This relation between metallicity and redshift was derived by \cite{Rafelski2012} based on the chemical abundance of a sample of damped Ly$\alpha{}$ systems up to redshift $\sim{}5$.

In {\emph{model~2}} we assume that all stars formed in a given redshift bin   $\Delta{}z$ have the same metallicity, but we calculate this metallicity as $\log{Z(z)/Z_{\odot}}=-0.19\,{}z$ if $z \leq 1.5$ and $\log{Z(z)/Z_{\odot}}=-0.22\,{}z$ if $z > 1.5$, which corresponds to rescaling the formula by \cite{Rafelski2012}, to obtain $Z(z=0)=Z_\odot$. In fact, the absolute metallicity calibration of the damped Ly$\alpha{}$ sample is quite uncertain, while the average metallicity of galaxies in the nearby Universe obtained from the Sloan Digital Sky Survey is consistent with the solar value \citep{Gallazzi2008, Madau2014}. In Table~\ref{tab:mergerrate}, we report our estimate of $R_{\rm loc, i}$ for all simulations and for these two models.

Figure~\ref{fig:rate} shows the local merger rate density of DNSs, BHNSs and BHBs for all simulations. The error bars account for the maximum difference between $N_{\rm cor,i}$ calculated for all metallicities. The uncertainty is relatively small (up to one order of magnitude) for DNSs, while it is very large (up to five orders of magnitude) for both BHNSs and BHBs, because  the number of BHNS and BHB mergers strongly depends on the progenitor's metallicity, while the number of DNS mergers is only mildly sensitive to progenitor's metallicity (Figure~\ref{fig:ncor}).

Our estimate of $R_{\rm loc, DNS}$ spans from $\sim{}5$ Gpc$^{-3}$ yr$^{-1}$ (for $\alpha{}1$) to $\sim{}10^3$ Gpc$^{-3}$ yr$^{-1}$ (for CC15$\alpha{}3$ and CC15$\alpha{}5$). The local merger rate density of BHNSs $R_{\rm loc, BHNS}$ spans from $\sim{}10^{-2}$ to $\sim{}10^3$ Gpc$^{-3}$ yr$^{-1}$, while the local merger rate density of BHBs $R_{\rm loc, BHB}$ spans from $\sim{}10^{-1}$ to $\sim{}4\times{}10^3$ Gpc$^{-3}$ yr$^{-1}$.

The simulations with high core-collapse SN kicks ($\alpha{}1$, $\alpha{}3$ and $\alpha{}5$) are not consistent with the local merger rate density of DNSs inferred from GW170817 \citep{Abbott2017d}. In contrast, the simulations with low core-collapse SN kicks and $\alpha{}\ge{}3$ (CC15$\alpha{}3$ and CC15$\alpha{}3$) are fairly consistent with the local merger rate of DNSs inferred from GW170817.  {\emph{Model~2}} applied to CC15$\alpha{}5$ gives an estimate of the local merger rate density of DNSs which is in good agreement with the value inferred from GW170817.

All simulations are consistent with the upper limit to the local merger rate density of BHNSs inferred from O1 data \citep{Abbott2016e}. {\emph{Model~1}} produces systematically larger rates than {\emph{model~2}}, because it includes more metal-poor stars by construction.

Finally, the uncertainty on the simulated BHB merger rate density is very large, much larger than the 90 per cent credible levels inferred from O1 data plus GW170104 \citep{Abbott2017a}. {\emph{Model~1}} predicts $R_{\rm loc,BHB}\gtrsim{}1000$ Gpc$^{-3}$ yr$^{-1}$, which is substantially larger than the 90 per cent credible levels inferred from current LIGO-Virgo results, whereas {\emph{model~2}} predicts  $R_{\rm loc,BHB}\sim{}40-90$ Gpc$^{-3}$ yr$^{-1}$, which is perfectly within the LIGO-Virgo 90 per cent credible interval.

In a companion paper \citep{Mapelli2018} we  model the metallicity evolution of the Universe by taking this information from a cosmological simulation. The local merger rate densities estimated by \cite{Mapelli2018} are consistent with the ones derived in this paper, within the uncertainties. 

Overall, we confirm the results of previous papers (e.g. \citealt{Chruslinska2018,Kruckow2018,Belczynski2018}): it is quite hard to match the local merger rate density of DNSs inferred from GW170817. Unlike \cite{Chruslinska2018}, who need to assume a different physics for CE evolution of DNS and BHB progenitors\footnote{\cite{Chruslinska2018} show that they can reproduce the merger rate of DNSs only if they allow some HG donors to survive a CE phase, but this assumption leads to a significant overestimate of the local BHB merger rate.}, we are able to match both the DNS and the BHB merger rates with the same physics, provided that low natal kicks are assumed for iron core-collapse SNe in binary systems. Natal kicks of core-collapse SNe are still matter of debate and addressing this open question is beyond the aims of this paper. Here, we simply show that assuming low kicks for iron core-collapse SNe in binary systems allows us to match the local merger rate of DNSs, BHNSs and BHBs all together.

\section{Conclusions}\label{sec:conclusions}
 We have investigated the formation of DNSs, BHNSs, BHBs from isolated binaries by means of our population-synthesis code {\sc MOBSE}. {\sc MOBSE} includes an updated formalism for mass loss by stellar winds, which depends on metallicity and Eddington ratio \citep{Giacobbo2018}, and incorporates several prescriptions for core-collapse SNe, electron-capture SNe, pulsational pair instability SNe and pair instability SNe. Here, we investigate the importance of CE ejection efficiency (assuming $\alpha{}=1,$ 3 and 5) and the impact of natal kicks of iron core-collapse SNe. In fact, the distribution of natal kicks is still matter of debate, and it might be that kicks in close binaries are lower than kicks in single stars (e.g. \citealt{Beniamini2016,Bray2016,Tauris2017}). Overall, we consider six sets of simulations: three with high core-collapse SN kicks (modeled through a Maxwellian curve with 1D rms 265 km s$^{-1}$, \citealt{Hobbs2005}) and with CE parameter $\alpha=1,$ 3 and 5 (simulations $\alpha{}1$, $\alpha{}3$ and $\alpha{}5$, see Table~\ref{tab:sims}) and three with low core-collapse SN kicks (modeled through a Maxwellian curve with 1D rms 15 km s$^{-1}$) and with CE parameter $\alpha=1,$ 3 and 5 (simulations CC15$\alpha{}1$, CC15$\alpha{}3$ and CC15$\alpha{}5$).

 We first look at the masses of DNSs, BHNSs and BHBs. The masses of NSs in DNSs span from $\sim{}1.1$ to $\sim{}2$ M$_\odot{}$, although low masses are more common than high masses, especially in merging DNSs formed with low natal kicks (simulations CC15$\alpha{}1$, CC15$\alpha{}3$ and CC15$\alpha{}5$, Figs~\ref{fig:mtot_DNS} and \ref{fig:m12DNS}). In contrast, the masses of NSs in  merging BHNSs tend to be preferentially high ($1.3-2.0$ M$_\odot$), while the masses of BHs in merging BHNSs tend to be preferentially low ($\sim{}5-15$ M$_\odot$), especially in simulations with low natal kicks  (simulations CC15$\alpha{}1$, CC15$\alpha{}3$ and CC15$\alpha{}5$, Figs~\ref{fig:mtot_BHNS} and \ref{fig:m12BHNS}).

 The masses of BHs in BHBs strongly depend on the progenitor's metallicity, more massive BHs being produced by metal-poor stars. Also, the number of BHBs depends on the progenitor's metallicity: metal-poor binaries tend to produce more BHBs than metal-rich ones. The maximum mass of BHs in simulated BHBs is $\sim{}65$ M$_\odot$, but the most massive BHBs do not merge within a Hubble time because their semi-major axes are too large. We find that the maximum mass of BHs in merging BHBs is $\sim{}45$ M$_\odot$ (Figs~\ref{fig:mtot_BHB} and \ref{fig:m12BHB}). This is consistent with the possible upper mass cut-off inferred from LIGO-Virgo data by \citet{Fishbach2017} and \citet{Wysocki2018}.

On the other hand, our binaries are evolved in isolation. If they were evolved in a dynamically active environment, such as a star cluster, some of the most massive BHs (with mass $45-65$ M$_\odot$) might still merge within a Hubble time, as a consequence of dynamical hardening or dynamical exchanges (e.g. \citealt{Mapelli2016}).
 
 We estimate the number of mergers per unit mass of the initial stellar population (Fig.~\ref{fig:ncor}). If core-collapse SN kicks are high (low), a DNS merger occurs every $\sim{}10^5-10^7$ M$_\odot$ ($\sim{}10^4-10^6$ M$_\odot$) of stellar population. Thus, low SN kicks boost the number of DNS mergers by at least a factor of 10. The number of DNS mergers per unit mass is weakly sensitive to progenitor's metallicity.

 In contrast, the number of both BHNS and BHB mergers per unit mass of stellar population depends on the progenitor's metallicity dramatically. A BHNS merger occurs every $\sim{}10^4-10^5$ M$_\odot$ of stellar population if $Z\leq{}0.01$ and every $\gtrsim{}10^7$ M$_\odot$ of stellar population if $Z>0.01$. The number of BHNS mergers per unit mass depends mildly on both the assumed CE parameters and  natal kicks. In particular, low kicks tend to produce more BHNS mergers at low metallicity and less at high metallicity.

 Similarly, a BHB merger occurs every $\sim{}10^4$ M$_\odot$ of stellar population if $Z\leq{}0.002$ and every $\sim{}10^7-10^8$ M$_\odot$ of stellar population if $Z\geq{}0.02$. The number of BHB mergers per unit mass depends only mildly on the CE parameters. There is no dependence of BHBs on the assumed natal kicks, but only because we assume that the kick is inversely proportional to the amount of fallback \citep{Fryer2012}. Thus, massive BHs receive always low kicks.

 Finally, we estimate the local merger rate density by convolving the number of mergers per unit stellar mass with the cosmic star formation rate density (equation~\ref{eq:rate}). Given the large uncertainties on the metallicity evolution of the Universe, we include the metallicity of the progenitor in our calculations to estimate the maximum uncertainty of our predicted merger rates. In all our simulations, the local merger rate density of BHNSs and BHBs is consistent with the values inferred from O1 LIGO-Virgo data, for reasonable assumptions about metallicity evolution. There is no significant dependence of the local merger rate of BHNSs and BHBs on CE parameters and on core-collapse SN kicks (Fig.~\ref{fig:rate}). 

In contrast, the local merger rate density of DNSs strongly depends on both CE parameters and SN kicks  (Fig.~\ref{fig:rate}). Only simulations with low SN kicks and high values of $\alpha$ (CC15$\alpha{}3$ and CC15$\alpha{}5$) match the local merger rate density inferred from GW170817 ($R_{\rm loc, DNS}=1540_{-1220}^{+3200}$ Gpc$^{-3}$ yr$^{-1}$, \citealt{Abbott2017d}). This result adds another piece to the intricate puzzle of natal kicks and DNS formation.
 
\section*{Acknowledgements}
We thank the anonymous referee for their comments which significantly improved this paper.
We thank Mario Spera, Abbas Askar, Chris Fryer, Daniel Wysocki and Alessandro Bressan for their useful comments. 
NG acknowledges financial support from Fondazione Ing. Aldo Gini and thanks the Institute for Astrophysics and Particle Physics of the University of Innsbruck for hosting him during the preparation of this paper. 
MM  acknowledges financial support from the MERAC Foundation through grant `The physics of gas and protoplanetary discs in the Galactic centre', from INAF through PRIN-SKA `Opening a new era in pulsars and compact objects science with MeerKat', from MIUR through Progetto Premiale 'FIGARO' (Fostering Italian Leadership in the Field of Gravitational Wave Astrophysics) and 'MITiC' (MIning The Cosmos 
Big Data and Innovative Italian Technology for Frontier Astrophysics and Cosmology), and from the Austrian National Science Foundation through FWF stand-alone grant P31154-N27 `Unraveling merging neutron stars and black hole - neutron star binaries with population-synthesis simulations'. 
 This work benefited from support by the International Space Science Institute (ISSI), Bern, Switzerland,  through its International Team programme ref. no. 393
{\it The Evolution of Rich Stellar Populations \& BH Binaries} (2017-18).
Numerical calculations have been performed through a CINECA-INFN agreement and through a CINECA-INAF agreement, providing access to resources on GALILEO and MARCONI at CINECA.



\bibliographystyle{mnras}
\bibliography{biblio} 






\bsp	
\label{lastpage}
\end{document}